\documentclass[aps,12pt,prb,amsmath,amssymb,floatfix,
superscriptaddress]{revtex4}

\usepackage{graphicx}

\def\be{\begin{equation}}       \def\ee{\end{equation}}
\def\bea{\begin{eqnarray}}      \def\eea{\end{eqnarray}}
\def\ba{\begin{array} }
\def\ea{\end{array} }
\def\bnum{\begin{enumerate} }
\def\enum{\end{enumerate}}

\def\=>{\Rightarrow}
\def\>{\rightarrow}

\def\eye2{Fathbb{I}}

\begin{document}
\title{Collective spin mode in a multi-component system of coupled itinerant
and localized electrons}
\author{Fan Yang}
\affiliation{Department of Physics, Beijing Institute of Technology, Beijing 100081,
P.R.China}
\author{Su-Peng Kou}
\affiliation{Department of Physics, Beijing Normal University, Beijing, 100875, P.R.China}
\author{Zheng-Yu Weng}
\email{weng@tsinghua.edu.cn}\affiliation{Institute for Advanced
Study , Tsinghua University, Beijing, 100084, P.R.China}

\begin{abstract}
We study collective spin excitations of a magnetically ordered state in a
multi-component system composed of both itinerant electrons and local
moments. Here the induced spin-density-wave (SDW) ordering of itinerant
electrons and the collinear antiferromangetic (AF) ordering of local moments
are locked together via a Hund's rule coupling. We show that the Goldstone
theorem still holds at the RPA level with the gapless spin wave protected
inside the small SDW gap of itinerant electrons, which, however, is fragile
in the presence of ion-anisotropy. A gapped \textquotedblleft
out-of-phase\textquotedblright \ spin mode extending over a much wider
energy scale above the SDW gap is found to be more robust against the
ion-anisotropy, which is mainly contributed by the local moment
fluctuations. While the scattering between the Goldstone mode and itinerant
electrons diminishes within the SDW gap, the \textquotedblleft
out-of-phase\textquotedblright \ mode will strongly interact with itinerant
electrons and thus dominate the spin and charge dynamics in such an ordered
phase. Possible relevance of such a model to the iron-pnictides will be also
discussed.
\end{abstract}
\date{\today}
\maketitle

\section{Introduction}

The recent discovery of high-$T_{c}$ superconductors in the iron pnictides
\cite{discovery} has renewed a tremendous interest in the interplay between
the magnetism and superconductivity. The similar issue has been vigorously
investigated in the high-$T_{c}$ cuprates over two decades, where
antiferromagnetism has been firmly related to the localized electrons of the
Mott insulator in undoped cuprate compounds\cite{rvb}. By contrast, in the
pnictides, the undoped parent material is not a simple Mott-Hubbard
insulator but resembles a multi-band bad metal - i.e., the iron 3$d$%
-electrons are believed to be quite itinerant with their hybridized
multi-orbitals forming multiple Fermi pockets at the Fermi level\cite%
{band4,band1,band2,band3,kuroki}.

It is natural for many to consider the spin-density-wave (SDW) order,
observed in the undoped parent compounds by the neutron scattering
measurements\cite{neutron}, as originated from the same itinerant electrons
via Fermi surface nesting \cite{band4}. This picture seems more consistent
in explaining the ARPES\cite{ludh,zhouxj,fengdl1}, transport\cite%
{discovery,transport}, and optical properties\cite{optical} than a purely
localized model, e.g., the \textrm{J}$_{1}$\textrm{-J}$_{2}$ model\cite%
{j1j20,j1j21,j1j22,j1j23}, where a Mott-insulator transition has been
implied. The latter is more reasonable in explaining the spin excitations in
neutron scattering \cite{j1j22} and high-temperature magnetic susceptibility%
\cite{susceptibility1,susceptibility2,gmzhang}, but obviously fails in
understanding the bad metal behavior\cite%
{badmetal1,discovery,optical,badmetal2} and the presence of the small Fermi
pockets\cite{ludh,zhouxj,fengdl1,dingh,kaminski} in the parent compounds.

Based on the overall experimental evidence, an alternative picture has been
recently proposed\cite{model}, which assumes that some kind of
orbital-selective Mott transition happens in the iron 3$d$ orbitals of the
pnictides such that both itinerant and Mott-localized electron coexist in
the system. The minimal model\cite{model} based on this picture tries to
reconcile the seemingly contradictory experimental facts and provides a
natural understanding of the unified driving force behind the collinear
antiferromangetic (AF) order and high-temperature superconductivity.
Recently a microscopic realization of an orbital-selective Mott transition
in the pnictides has been studied\cite{dmft1,dmft} based on the dynamic
mean-field theory, which lends further support to this model.

The key and unique feature for such a coexistent itinerant and localized
electron system is that the two subsystems share the \emph{same}
characteristic momenta at $\mathbf{Q}_{s}=$ $(\pi ,0)$ or $(0,\pi )$.
Namely, the hole and electron Fermi surface pockets around $\Gamma $ and $M$
points in the Brillouin zone (BZ) are approximately connected by $\mathbf{Q}%
_{s}$ in the undoped case (i.e., close to the Fermi surface nesting), and at
the same time, the local moments are strongly correlated at the AF
wavevectors $\mathbf{Q}_{s}$. As a consequence, the local Hund's rule
coupling between the itinerant electrons and local moments can be
significantly enhanced around $\mathbf{Q}_{s}$, which is called the
\textquotedblleft \emph{resonant effect\textquotedblright }\cite{model}\emph{%
.} At low temperature, such a \textquotedblleft resonant
effect\textquotedblright \ can serve as a predominant force in driving the
\emph{magnetic} or \emph{pairing instability} at different dopings. Here the
magnetic phase is predicted\cite{model} to be an induced SDW order of the
itinerant electrons locking with the collinear AF order of the local moments
at the \emph{same} $\mathbf{Q}_{s}$. Corresponding to such an AF ordering, a
small SDW gap will open up in the excitation spectrum of itinerant
electrons, although \emph{not }necessarily pinned at the Fermi level as in
an SDW state purely driven by Fermi surface nesting. At the mean-field
level, the low-lying AF fluctuation of the local moments is also gapped at $%
\mathbf{Q}_{s}$ due to the mutual locking of the magnetic orders in the two
subsystems. Therefore, after the spontaneous magnetic symmetry breaking, the
strong \textquotedblleft resonant\textquotedblright \ scattering between the
two degrees of freedom gets substantially reduced, which leads to a very
coherent charge transport contributed by the ungapped part of the Fermi
surfaces in consistency with the optical measurement\cite{optical}.

However, it remains an important issue whether a gapless spin wave, i.e., a
Goldstone mode, is still present in the magnetically ordered state of such a
multi-component system. To answer this question, one has to go beyond the
mean-field theory to study the collective spin fluctuations, which is also
important in order to self-consistently address the issue how the charge
dynamics gets reshaped in the AF phase. In this paper, we shall address this
issue with using a realistic five-band model\cite{kuroki} to characterize
the itinerant electrons near the Fermi pockets, and a \textrm{J}$_{1}$%
\textrm{-J}$_{2}$ type model to describe the Mott-localized
electrons. Then we study the Hund's rule coupling between the
itinerant and localized electrons at the RPA level in the magnetic
ordered phase. We demonstrate that the Goldstone theorem indeed
holds at the RPA level as a gapless spin wave emerges within the
mean-field SDW gap of itinerant electrons. But it is fragile
against the ion-anisotropy. We further find that the coupling
between the Goldstone mode and the charge carriers diminishes in
the long-wavelength around $\mathbf{Q}_{s}$\ as expected. On the
other hand, distinct from a single component system, a gapped
\textquotedblleft out-of-phase\textquotedblright \ collective spin
mode is also present with its high-energy part predominantly
contributed by the local moments whose energy scale $\sim J_{2}$
extends over a much wider regime than the SDW gap. Its low-energy
part gets strongly renormalized by coupling to the itinerant
electrons around $\mathbf{Q}_{s}$---it becomes gapped once the
mean-field SDW order forms by itinerant electrons, which is not
significantly modified at the RPA level. Furthermore, such an
\textquotedblleft out-of-phase\textquotedblright \ mode is not
sensitive to a weak ion-anisotropy effect and is thus more robust
than the Goldstone mode. The existence of the two branches of spin
excitations in the AF state is a unique prediction of the present
multi-component model. In particular, it is this \textquotedblleft
out-of-phase\textquotedblright \ spin mode that remains strongly
interacting with itinerant electrons, at an energy higher than its
gap, and therefore dominates the high-energy magnetic and
transport properties in the magnetically ordered phase.

The remainder of the paper is organized as follows. In Sec. II, we introduce
the model and present the mean-field treatment in the magnetically ordered
state. Then in Sec. III, we discuss the spin dynamics at the RPA level and
demonstrate that the spin collective excitations are split into a gapless
Goldstone mode which is upper-bounded and a gapped out-of-phase mode which
extends over a much wider energy scale. In Sec. IV, we study the scattering
between the collective spin modes and the itinerant electrons based on the
single-particle self-energy of the itinerant electrons and the optical
conductivity, which illustrate that the out-of-phase mode will play a
dominant role beyond its energy gap. Finally, Sec. V is devoted to the
discussion and conclusion.

\section{A multi-component system of coupled itinerant and localized
electrons}

\subsection{Model}

We consider a multi-component system composed of coexistent multiband
itinerant and Mott-localized electrons described by
\begin{equation}
H=H_{\mathrm{it}}+H_{\mathrm{lo}}+H_{J_{H}}.  \label{H}
\end{equation}

Here $H_{\mathrm{it}}$ is a tight-binding model of multiband itinerant
electrons:
\begin{equation}
H_{\mathrm{it}}=-\sum_{i,j,m,n,\sigma }t_{ij,mn}c_{im\sigma }^{\dagger
}c_{jn\sigma },  \label{hitr}
\end{equation}%
where $m$ and $n$ are the orbital indices. The hopping integral $t_{ij,mn}$
in $H_{\mathrm{it}}$ will be given based on a realistic five-band
tight-binding model proposed\cite{kuroki} to describe the undoped
iron-pnictide materials. The resulting band structure near the Fermi energy
is shown in Fig. 1 by the solid (black) curves, with the Fermi surface shown
in Fig. 2(a) in the undoped case (i.e., six electrons per site). Note that
some slight modification with a global renormalization factor reducing the
bandwidth has been phenomenologically made here in order to be consistent
with ARPES\cite{ludh,zhouxj,fengdl1,kaminski}. As shown in Fig. 2(a), the
itinerant electrons form hole and electron pockets at the Fermi energy,
which are located at the $\Gamma $ point and the $M$ point, respectively,
separated by the momenta $\mathbf{Q}_{s}$ in an extended BZ.

The second term $H_{\mathrm{lo}}$ in (\ref{H}) describes the localized
electrons in which a Mott gap is opened up via the so-called
orbital-selective Mott transition\cite{model}. Namely, the corresponding
electrons only contribute to spin fluctuations, near the Fermi energy, by
the local moments formed from the filled lower Hubbard band. Note that a
microscopic realization of the orbital-selective Mott transition in such a
system has been recently discussed based on the dynamic mean-field theory%
\cite{dmft1,dmft}. We shall simply use a $J_{1}$-$J_{2}$ model of $S=1$ to
depict the superexchange couplings between these local moments, i.e.,
\begin{equation}
H_{\mathrm{lo}}=J_{1}\sum_{\left \langle ij\right \rangle }\mathbf{\hat{S}}%
_{i}\cdot \mathbf{\hat{S}}_{j}+J_{2}\sum_{\left \langle \left \langle
ij\right \rangle \right \rangle }\mathbf{\hat{S}}_{i}\cdot \mathbf{\hat{S}}%
_{j}  \label{hlo}
\end{equation}%
where $\left \langle ij\right \rangle $ and $\left \langle \left \langle
ij\right \rangle \right \rangle $ denote the nearest-neighbor and
next-nearest-neighbor coupling, respectively. Here we assume $J_{1}<2J_{2}$
such that the ground state of $H_{\mathrm{lo}}$ itself may form a collinear
AF ordering at the wavevector $\mathbf{Q}_{s}$. In principle, the same
five-band electrons in iron-pnictides should contribute to both itinerant
and local moment degrees of freedom. But for simplicity we treat $H_{\mathrm{%
it}}$ and $H_{\mathrm{lo}}$ as if they govern\emph{\ independent} degrees of
freedom in the low-energy sector near the Fermi energy, so long as the Mott
gap remains a large energy scale. We ignore the issues like how the Fermi
surface shape gets affected by the orbital-selective Mott transition as well
as how the Luttinger volume is correctly accounted for and mainly focus on
the low-energy physics in the present work.

The third term $H_{J_{H}}$ in (\ref{H}) is the Hund's rule coupling between
the spins of itinerant and the localized electrons:

\begin{equation}
H_{J_{H}}=-\sum_{i,m}J_{0}^{m}\mathbf{\hat{S}}_{i}\cdot \mathbf{\hat{s}}%
_{im}.  \label{hjh}
\end{equation}%
where $\mathbf{\hat{s}}_{im}=c_{im}^{\dagger }\mathbf{\hat{\sigma}}c_{im}$
is the spin operator of the itinerant electrons in the m-th orbital and $%
\mathbf{\hat{S}}_{i}$ denotes the localized moment at site $i$. $J_{0}^{m}$
is a renormalized Hund's rule coupling constant. For simplicity, we shall
assume a single $J_{0}^{m}=J_{0}$ for different orbitals throughout the
paper.

Originally a simpler form of (\ref{H}) was proposed as a minimal model to
describe the low-energy physics in the iron-pnictides\cite{model}. The most
important feature in such a model Hamiltonian is that the peculiar momenta $%
\mathbf{Q}_{s}$, which on the one hand connects the two Fermi pockets of the
itinerant electrons at $\Gamma $ and $M$ points, respectively, and on the
other hand coincide with the AF wavevectors of the local moments. It implies
\emph{strongly enhanced} dynamic coupling between the itinerant and local
moment degrees of freedom, once the short-range AF correlations of the local
moments set in around $\mathbf{Q}_{s}$ even in a high-temperature normal
state. With the decrease of temperature, such a \textquotedblleft
resonant\textquotedblright \ coupling will result in an AF ordered phase
with distinctive dynamic behaviors to be explored below, as compared to an
ordinary SDW state of a pure itinerant electron system due to the
Fermi-surface nesting mechanism or the collinear AF state of a pure $J_{1}$-$%
J_{2}$ mode.

\begin{figure}[tbp]
\includegraphics[clip,width=0.7\textwidth]{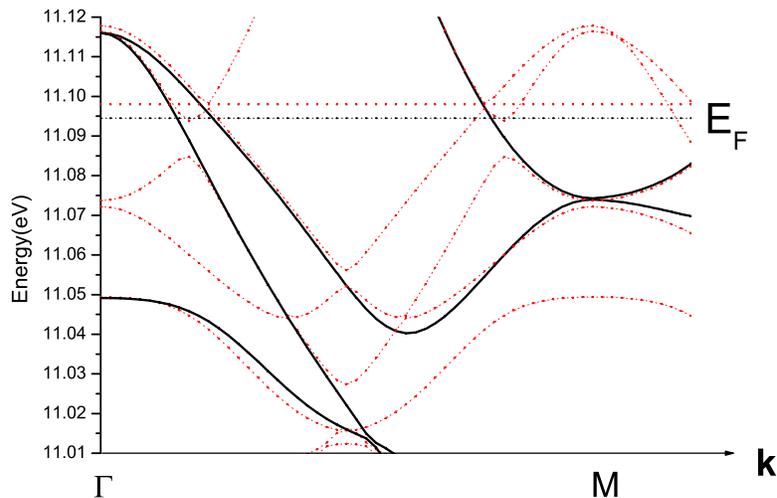}
\caption{(Color online) The band structure near the Fermi energy for
the five-band model described by $H_{\mathrm{it}}$ in
(\protect\ref{hitr}) (black solid curves) in the undoped case. The
reconstruction of the band structure of itinerant electrons in the
presence of a collinear AF order is shown by red dotted curves.}
\label{Fig.1}
\end{figure}

\begin{figure}[tbp]
\includegraphics[clip,width=0.6\textwidth]{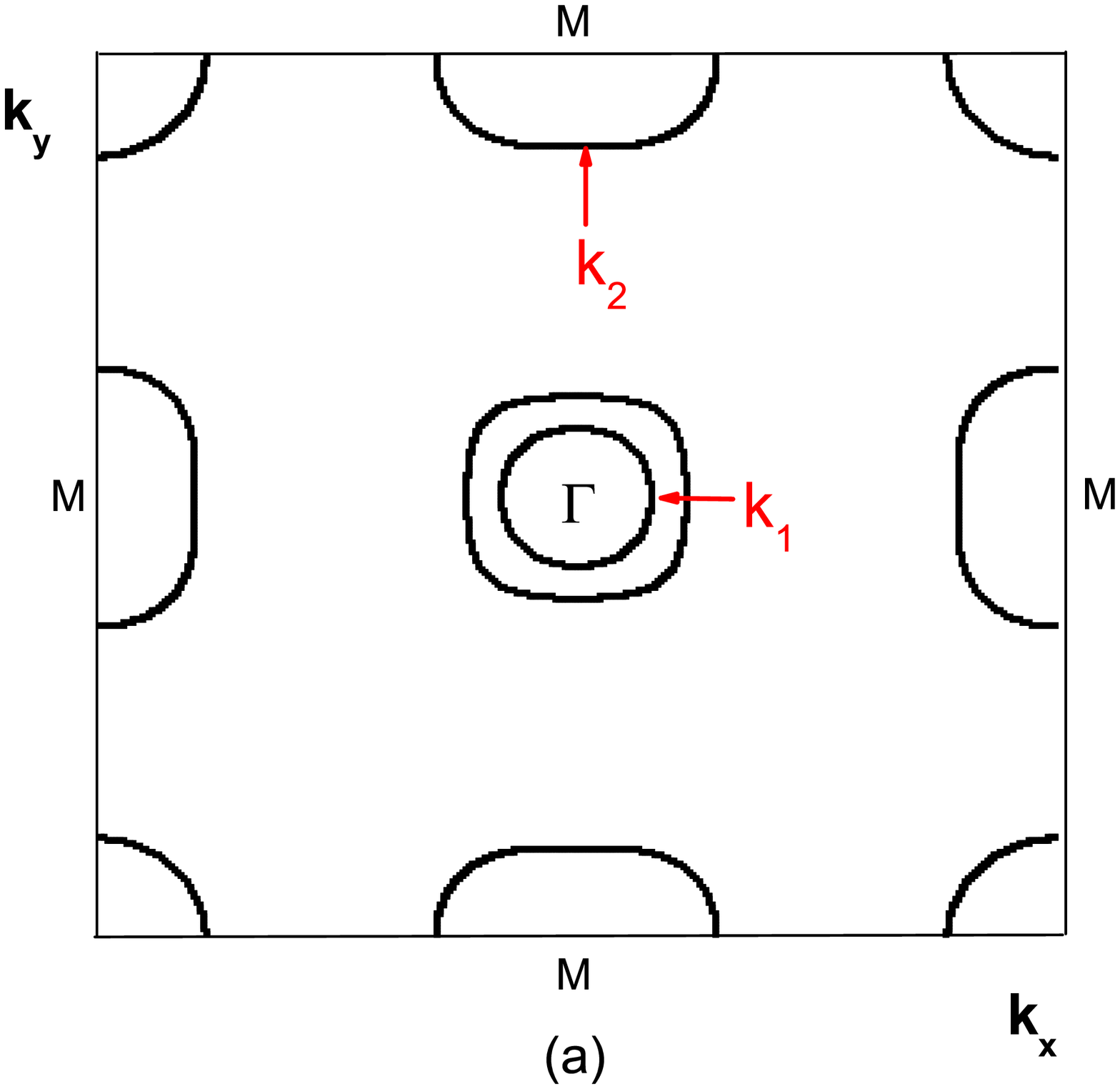} %
\includegraphics[clip,width=0.5\textwidth]{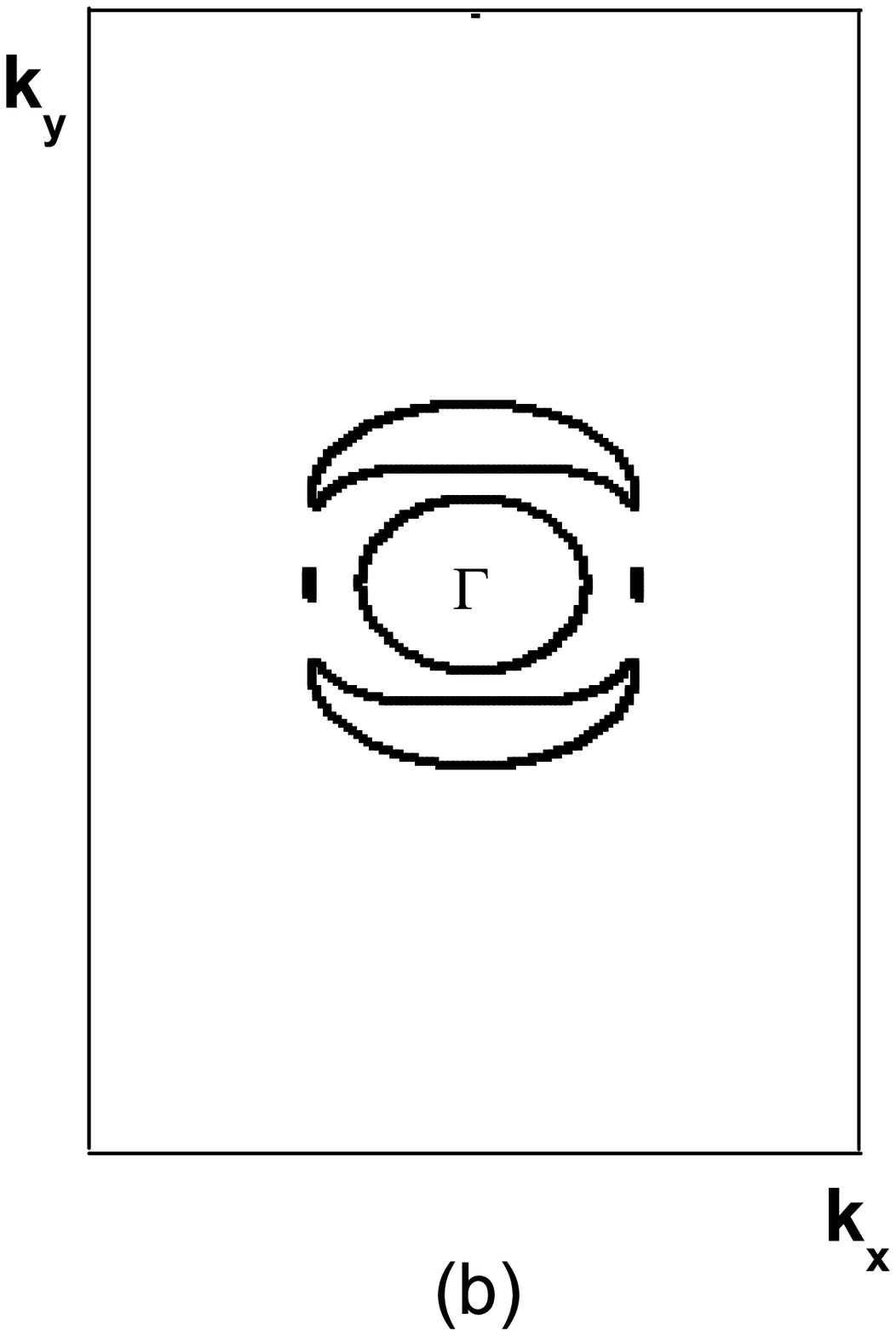}
\caption{(Color online) (a): The Fermi pockets of the itinerant electrons in
the undoped case. (b): The reconstruction of the Fermi surface in the
presence of a collinear AF order with a wavevector $\mathbf{Q}_{s}=(\protect%
\pi ,0)$ in the reduced BZ. }
\label{Fig.2}
\end{figure}

\subsection{Mean-field treatment in the collinear AF ordered state}

In the following we first use a mean-field approximation to study the
collinear AF ordered state in (\ref{H}). The spin and charge dynamics at the
RPA level will be investigated in the later sections.

We start with the interaction term $H_{J_{H}}$ in (\ref{hjh}) between the
two sub-systems. By introducing two order parameters of magnetization for
the local-moments, $M_{\left( \mathrm{lo}\right) }$, and
itinerant-electrons, $M_{\left( \mathrm{it}\right) }$, respectively, as in $%
\left \langle \hat{S}_{i}^{z}\right \rangle \equiv M_{\left( \mathrm{lo}%
\right) }e^{i\mathbf{Q}_{s}\cdot \mathbf{r}_{i}}$ and $\sum_{m}\left \langle
\hat{s}_{im}^{z}\right \rangle \equiv M_{\left( \mathrm{it}\right) }e^{i%
\mathbf{Q}_{s}\cdot \mathbf{r}_{i}}$, one obtains the following
linearization in $H_{J_{H}}$, given by

\begin{eqnarray}
H_{J_{H}} &\rightarrow &H_{I}=-J_{0}\sum_{i}\left[ M_{\left( \mathrm{lo}%
\right) }e^{i\mathbf{Q}_{s}\cdot \mathbf{r}_{i}}\sum_{m}s_{im}^{z}+M_{\left(
\mathrm{it}\right) }e^{i\mathbf{Q}_{s}\cdot \mathbf{r}_{i}}\hat{S}_{i}^{z}%
\right]  \nonumber \\
&\equiv &H_{I\left( \mathrm{it}\right) }+H_{I\left( \mathrm{lo}\right) }.
\label{couple}
\end{eqnarray}%
Below the effect of such mean-field terms on the local moments and itinerant
electrons will be explored in a self-consistent way.

\subsubsection{Local moment part}

In the magnetic order phase with $\left \langle \hat{S}_{i}^{z}\right
\rangle =M_{\left( \mathrm{lo}\right) }e^{i\mathbf{Q}_{s}\cdot \mathbf{r}%
_{i}}$, one may first use the conventional spin-wave approximation to treat $%
H_{\mathrm{lo}}$ in (\ref{hlo}) and then add $H_{I\left( \mathrm{lo}\right)
} $ in (\ref{couple}) to incorporate the effect of the coupling to the
itinerant electrons. Here $\mathbf{Q}_{s}$ is chosen to be $%
\left(\pi,0\right)$.

Introduce the Holstein-Primakoff (HP) transformation
\begin{eqnarray}
\hat{S}_{iA}^{+} &=&\sqrt{2S-a_{i}^{\dagger }a_{i}}a_{i},\text{ }\hat{S}%
_{iA}^{-}=a_{i}^{\dagger }\sqrt{2S-a_{i}^{\dagger }a_{i}},\text{ }\hat{S}%
_{iA}^{z}=S-a_{i}^{\dagger }a_{i},  \nonumber \\
\hat{S}_{jB}^{+} &=&\sqrt{2S-b_{j}^{\dagger }b_{j}}b_{j}^{\dagger },\text{ }%
\hat{S}_{jB}^{-}=b_{j}\sqrt{2S-b_{j}^{\dagger }b_{j}},\text{ }\hat{S}%
_{jB}^{z}=-S+b_{j}^{\dagger }b_{j},  \label{hptr}
\end{eqnarray}%
where $A$ and $B$ sublattices are defined by the staggered factor $e^{i%
\mathbf{Q}_{s}\cdot \mathbf{r}_{i}}=\pm 1$. Under the approximation $\sqrt{%
2S-a_{i}^{\dagger }a_{i}}\approx $ $\sqrt{2S-b_{i}^{\dagger }b_{i}}\approx
\sqrt{2S}$ and using the boson operators in the momentum space
\begin{eqnarray}
a_{i} &=&\left( \frac{2}{N}\right) ^{1/2}\sum \nolimits_{\mathbf{k}}^{\prime
}a_{\mathbf{k}}\exp (i\mathbf{k}\cdot \mathbf{r}_{i}), \\
b_{j} &=&\left( \frac{2}{N}\right) ^{1/2}\sum \nolimits_{\mathbf{k}}^{\prime
}b_{\mathbf{k}}\exp (i\mathbf{k}\cdot \mathbf{r}_{j}),
\end{eqnarray}%
$H_{\mathrm{lo}}$ is transformed into
\begin{equation}
H_{\mathrm{lo}}=S\sum \nolimits_{\mathbf{k}}^{\prime }[\Gamma _{\mathbf{k}%
}(a_{\mathbf{k}}^{\dagger }a_{\mathbf{k}}+b_{\mathbf{k}}^{\dagger }b_{%
\mathbf{k}})+M_{\mathbf{k}}(a_{\mathbf{k}}b_{-\mathbf{k}}+a_{\mathbf{k}%
}^{\dagger }b_{-\mathbf{k}}^{\dagger })],  \label{HSW}
\end{equation}%
where $\sum_{\mathbf{k}}^{^{\prime }}$ means that the sum is within a
reduced BZ with
\begin{eqnarray}
\Gamma _{\mathbf{k}} &=&4J_{2}+2J_{1}\cos k_{y},  \label{gamma} \\
M_{\mathbf{k}} &=&2J_{1}\cos k_{x}+4J_{2}\cos k_{x}\cdot \cos k_{y}.
\label{M}
\end{eqnarray}%
Then (\ref{HSW}) can be diagonalized by the Bogolubov transformation
\begin{eqnarray}
a_{\mathbf{k}} &=&u_{\mathbf{k}}\alpha _{k}+v_{\mathbf{k}}\beta _{-\mathbf{k}%
}^{\dagger }  \nonumber \\
b_{-\mathbf{k}}^{+} &=&v_{\mathbf{k}}\alpha _{k}+u_{\mathbf{k}}\beta _{-%
\mathbf{k}}^{\dagger },  \label{bgtr2}
\end{eqnarray}%
as follows
\begin{equation}
H_{\mathrm{lo}}=\sum \nolimits_{\mathbf{k}}^{\prime }\omega _{\mathbf{k}%
}(\alpha _{\mathbf{k}}^{\dagger }\alpha _{\mathbf{k}}+\beta _{\mathbf{k}%
}^{\dagger }\beta _{\mathbf{k}}),  \label{HSWD}
\end{equation}%
where
\begin{eqnarray}
u_{\mathbf{k}} &=&\left( \frac{\omega _{\mathbf{k}}+\Gamma _{\mathbf{k}}}{%
2\omega _{\mathbf{k}}}\right) ^{\frac{1}{2}},  \label{bgtr1} \\
v_{\mathbf{k}} &=&-\left( \frac{-\omega _{\mathbf{k}}+\Gamma _{\mathbf{k}}}{%
2\omega _{\mathbf{k}}}\right) ^{\frac{1}{2}}\mathrm{sgn}(M_{\mathbf{k}}),
\label{bg2} \\
\omega _{\mathbf{k}} &=&S\sqrt{\Gamma _{\mathbf{k}}^{2}-M_{\mathbf{k}}^{2}},
\label{bgtr3}
\end{eqnarray}%
Here $\omega _{\mathbf{k}}$ is gapless at $\mathbf{k=Q}_{s}$ as the
Goldstone mode of $H_{\mathrm{lo}}$ in the AF ordered phase.

Now let us add $H_{I\left( \mathrm{lo}\right) }$ in (\ref{couple}) arising
from the mean-field decoupling of $H_{J_{H}}$. It can be reexpressed in the
spin-wave formalism by%
\begin{equation}
H_{ I\left( \mathrm{lo}\right)}=J_{0}M_{\left( \mathrm{it}\right) }\sum
\nolimits_{\mathbf{k}}^{\prime }\left( a_{\mathbf{k}}^{\dagger }a_{\mathbf{k}%
}+b_{\mathbf{k}}^{\dagger }b_{\mathbf{k}}\right) +\mathrm{const}.
\label{couplelosp}
\end{equation}%
This term will lead to a shift in $\Gamma _{\mathbf{k}}$ defined in (\ref%
{gamma}): i.e., $\Gamma _{\mathbf{k}}\rightarrow \Gamma _{\mathbf{k}%
}+J_{0}M_{\left( \mathrm{it}\right) }/S$. As a result, the dispersion of the
spin wave is modified by
\begin{equation}
\omega _{\mathbf{k}}\rightarrow \omega _{\mathbf{k}}=S\sqrt{\left( \Gamma _{%
\mathbf{k}}+J_{0}M_{\left( \mathrm{it}\right) }/S\right) ^{2}-M_{\mathbf{k}%
}^{2}}.  \label{dispersionspinwavemeanfield}
\end{equation}%
In particular, $\omega _{\mathbf{k}}$ is no longer gapless at $\mathbf{Q}%
_{s} $ with\ an energy gap induced by $M_{\left( \mathrm{it}\right) }$ as
\begin{equation}
\omega _{\mathbf{Q}_{s}}=\sqrt{4M_{\left( \mathrm{it}\right) }SJ_{0}\left(
2J_{2}+J_{1}\right) +\left( M_{\left( \mathrm{it}\right) }J_{0}\right) ^{2}}.
\label{logap}
\end{equation}

\begin{figure}[tbp]
\includegraphics[clip,width=0.7\textwidth]{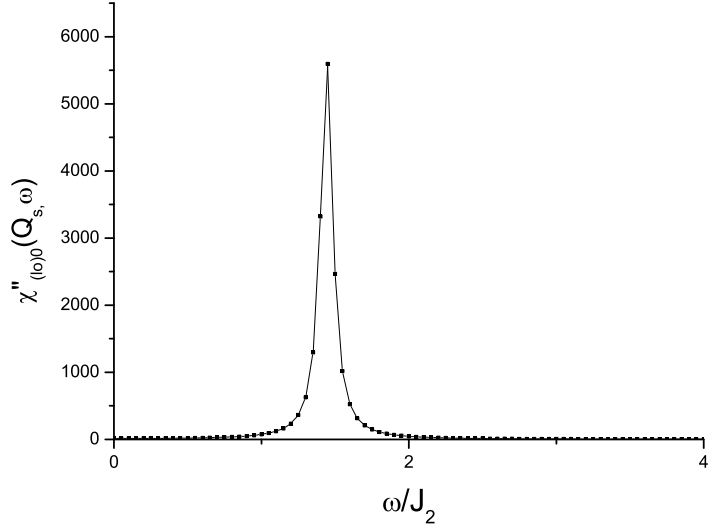}
\caption{(Color online) Dynamic spin susceptibility of local-moments at
mean-field level, i.e. $\protect\chi^{\prime \prime}_{\left(\mathrm{lo}%
\right)0\mathbf{q}}\left(\protect\omega \right)\equiv -2\mathrm{Im}\protect%
\chi^{+-}_{\left(\mathrm{lo}\right)0}\left(\mathbf{q},\mathbf{q},\protect%
\omega \right)$ at $\mathbf{q}=\mathbf{Q}_{s}$ shows a gap, given by (%
\protect\ref{logap}), which is opened up in the presence of a mean-field
coupling to the SDW ordering of itinerant electrons, with the parameters
given in Sec. IIB.}
\label{Fig3}
\end{figure}
The order parameter $M_{\left( \mathrm{lo}\right) }$ can be
self-consistently calculated as
\begin{eqnarray}
M_{\left( \mathrm{lo}\right) } &=&S-\frac{1}{N}\left \langle \sum \nolimits_{%
\mathbf{k}}^{\prime }\left( a_{\mathbf{k}}^{\dagger }a_{\mathbf{k}}+b_{%
\mathbf{k}}^{\dagger }b_{\mathbf{k}}\right) \right \rangle  \nonumber \\
&=&S-\frac{2}{N}\sum \nolimits_{\mathbf{k}}^{\prime }v_{\mathbf{k}}^{2},
\label{SELFCON1}
\end{eqnarray}%
where $v_{\mathbf{k}}$ is defined by (\ref{bg2}) with $\Gamma _{\mathbf{k}%
}\rightarrow \Gamma _{\mathbf{k}}+J_{0}M_{\left( \mathrm{it}\right) }/S$ and
$\omega _{\mathbf{k}}$ defined in (\ref{dispersionspinwavemeanfield}).

Finally, the spin-spin correlations for the local-moment defined by
\[
\chi _{\left( \mathrm{lo}\right) }^{+-}(\mathbf{q},\mathbf{q}%
,t)=-i\left
\langle T\hat{S}_{\mathbf{q}}^{+}(t)\hat{S}_{-\mathbf{q}%
}^{-}(0)\right
\rangle ,\text{ }\chi _{\left( \mathrm{lo}\right) }^{+-}(%
\mathbf{q},\mathbf{q}+\mathbf{Q}_{s},t)=-i\left \langle T\hat{S}_{\mathbf{q}%
}^{+}(t)\hat{S}_{-\mathbf{q}-\mathbf{Q}_{s}}^{-}(0)\right \rangle ,
\]%
can be obtained for this mean-field spin-wave state in the frequency space
as
\begin{eqnarray}
\chi _{\left( \mathrm{lo}\right) 0}^{+-}(\mathbf{q},\mathbf{q},\omega )
&=&M_{\left( \mathrm{lo}\right) }\left( u_{\mathbf{k}}+v_{\mathbf{k}}\right)
^{2}\left( \frac{1}{\omega -\omega _{\mathbf{k}}+i0^{+}}+\frac{1}{-\omega
-\omega _{\mathbf{k}}+i0^{+}}\right) ,  \nonumber \\
\chi _{\left( \mathrm{lo}\right) 0}^{+-}(\mathbf{q},\mathbf{q}+\mathbf{Q}%
_{s},\omega ) &=&M_{\left( \mathrm{lo}\right) }\left( \frac{1}{\omega
-\omega _{\mathbf{k}}+i0^{+}}-\frac{1}{-\omega -\omega _{\mathbf{k}}+i0^{+}}%
\right) ,  \label{SPGR1}
\end{eqnarray}%
The spin gap in (\ref{logap}) is clearly illustrated in the dynamic spin
susceptibility $\chi _{\left( \mathrm{lo}\right) 0\mathbf{q}}^{\prime \prime
}\left( \omega \right) \equiv -2\mathrm{Im}\chi _{\left( \mathrm{lo}\right)
0}^{+-}\left( \mathbf{q},\mathbf{q},\omega \right) $ at $\mathbf{q}=\mathbf{Q%
}_{s}$ as shown in Fig. \ref{Fig3} (the parameters used are to be given
below). It is pointed out that in the above calculation, we have further
used the approximation: $S_{iA\left( B\right) }^{+}\rightarrow \sqrt{%
2M_{\left( \mathrm{lo}\right) }}a_{i}^{\dagger }$ $\left( \sqrt{2M_{\left(
\mathrm{lo}\right) }}b_{i}^{\dagger }\right) $ in the original HP
transformation (\ref{hptr}) such that the spin commutation relations are
satisfied at the mean-field level, i.e. $\left \langle \mathbf{\hat{S}}%
\times \mathbf{\hat{S}}\right \rangle =i\hbar \left \langle \mathbf{\hat{S}}%
\right \rangle $.

\subsubsection{Itinerant electron part}

Combining $H_{I\left( \mathrm{it}\right) }$ in (\ref{couple}) with the band
kinetics energy term $H_{\mathrm{it}}$ in (\ref{hitr}), the mean-field
Hamiltonian of the itinerant-electrons reads
\begin{eqnarray}
H_{\mathrm{it}}+H_{I\left( \mathrm{it}\right) } &=&\sum \nolimits_{\mathbf{k}%
mn\sigma }^{\prime }\left[f_{mn}(\mathbf{k})c_{\mathbf{k}m\sigma }^{+}c_{%
\mathbf{k}n\sigma }+f_{mn}(\mathbf{k}+\mathbf{Q}_{s})c_{\mathbf{k}+\mathbf{Q}%
_{s},m\sigma }^{\dagger }c_{\mathbf{k}+\mathbf{Q}_{s},n\sigma }\right.
\nonumber \\
&&-\left. \frac{J_{0}M_{\left( \mathrm{lo}\right) }\sigma }{2}\delta
_{mn}\left(c_{\mathbf{k}+\mathbf{Q}_{s},m\sigma }^{\dagger }c_{\mathbf{k}%
,n\sigma }+c_{\mathbf{k},m\sigma }^{\dagger }c_{\mathbf{k}+\mathbf{Q}%
_{s},n\sigma }\right)\right]  \nonumber \\
&\equiv &\sum \nolimits_{\mathbf{k}\sigma }^{\prime }X_{\mathbf{k}\sigma
}^{\dagger }H_{\sigma }X_{\mathbf{k}\sigma },  \label{sdwhit}
\end{eqnarray}%
where
\begin{equation}
f_{mn}\left( \mathbf{k}\right) =2\sum_{\mathbf{r}_{i}-\mathbf{r}%
_{j}}t_{ij,mn}e^{i\mathbf{k}\cdot \left( \mathbf{r}_{i}-\mathbf{r}%
_{j}\right) }.  \label{fmn}
\end{equation}%
In the second line of (\ref{sdwhit}), the $10\times 10$ matrix $H_{\sigma }$
for the five bands is defined by%
\begin{equation}
\left( H\right) _{\sigma }=\left(
\begin{array}{cc}
F_{\mathbf{k}} & -\gamma I\sigma \\
-\gamma I\sigma & F_{\mathbf{k}+\mathbf{Q}_{s}}%
\end{array}%
\right) ,  \label{10by10matrix}
\end{equation}%
with
\begin{equation}
\gamma =\frac{J_{0}M_{\left( \mathrm{lo}\right) }}{2}.  \label{alpha}
\end{equation}%
Here $I$ is the $5\times 5$ identity matrix and $F$ is the matrix defined by
$\left( F\right) _{m,n}=f_{mn}$. The column vector $X_{\mathbf{k}\sigma }$
is given by
\begin{equation}
X_{\mathbf{k}\sigma }^{T}\equiv \left( c_{\mathbf{k},1\sigma},c_{\mathbf{k}%
,2\sigma}\cdots c_{\mathbf{k}+\mathbf{Q}_{s},1\sigma }c_{\mathbf{k}+\mathbf{Q%
}_{s},2\sigma }\cdots \right) .  \label{coperator}
\end{equation}

By diagonalizing the $10\times 10$ matrix $\left( H\right) _{\sigma }$
\begin{equation}
U_{\sigma }^{\dagger }H_{\sigma }U_{\sigma }=D,  \label{diagonal}
\end{equation}%
one gets
\begin{equation}
H_{\mathrm{it}}+H_{I\left( \mathrm{it}\right) }=\sum \nolimits_{\mathbf{k}%
\alpha \sigma }^{\prime }E_{\mathbf{k}\alpha }c_{\mathbf{k}\alpha \sigma
}^{\dagger }c_{\mathbf{k}\alpha \sigma },  \label{band2}
\end{equation}%
where the band energy $E_{\mathbf{k}\alpha }$ is equal to the $\alpha $-th
diagonal element of $D$, presented in Fig. 1 by the red dotted curves with
the corresponding Fermi surface in the reduced BZ shown in Fig. 2(b) in the
undoped case. Here the order parameter $M_{\left( \mathrm{it}\right) }$ is
self-consistently determined by
\begin{eqnarray}
M_{\left( \mathrm{it}\right) } &=&\frac{1}{2N}\sum_{in}e^{i\mathbf{Q}%
_{s}\cdot \mathbf{r}_{i}}\left \langle c_{in\uparrow }^{\dagger
}c_{in\uparrow }-c_{in\downarrow }^{\dagger }c_{in\downarrow }\right \rangle
\nonumber \\
&=&\frac{2}{N}\sum \nolimits_{E_{\mathbf{k}\alpha }<E_{F}}^{\prime
}\sum_{n}U_{\mathbf{k}\uparrow }^{\ast }\left( n+5,\alpha \right) U_{\mathbf{%
k}\uparrow }\left( n,\alpha \right) .  \label{selfcon2}
\end{eqnarray}%
Together with (\ref{SELFCON1}), we find $M_{\left( \mathrm{it}\right)
}=0.252 $ and $M_{\left( \mathrm{lo}\right) }=0.892$ by choosing $J_{0}=20$ $%
\mathrm{meV}$ at $J_{1}=0,$ and $J_{2}=20$ $\mathrm{meV}$ for $S=1$. In the
following we shall use these parameters to examine various spin and charge
dynamics (we have also checked other small ratios of $J_{1}/J_{2}$ and found
the results remain qualitatively unchanged).

Similar mean-field results have been previously obtained\cite{model} in a
simpler model of (\ref{H}), where a spin gap similar to (\ref{logap}) is
also found in the spin-wave spectrum of local moments due to the Hund's rule
coupling to the SDW order of the itinerant electrons. Such a gap will
protect the collinear AF ordering jointly formed by \emph{both} local
moments and itinerant electrons below a transition temperature $T_{\mathrm{%
SDW}}.$ For example, the presence of this gap is reflected by a steep
reduction of the uniform spin susceptibility below $T_{\mathrm{SDW}}$, which
is consistent\cite{model} with the experimental measurement in the
iron-pnictides\cite{susceptibility1,susceptibility2}. However, according to
the Goldstone theorem, a gapless mode is generally expected to exist in the
AF ordered state. How such a Goldstone mode can be reconciled with the
gapped local moment fluctuations discussed above will be studied at the RPA
level in the next section.

Finally, the spin-spin correlation functions of itinerant electrons are
defined as
\[
\chi _{\left( \mathrm{it}\right) }^{+-}(\mathbf{q},\mathbf{q,}%
t)=-i\left
\langle T\hat{s}_{\mathbf{q}}^{+}(t)\hat{s}_{-\mathbf{q}%
}^{-}(0)\right
\rangle ,\text{ \ \ }\chi _{\left( \mathrm{it}\right) }^{+-}(%
\mathbf{q},\mathbf{q+\mathbf{Q}_{s},}t)=-i\left \langle T\hat{s}_{\mathbf{q}%
}^{+}(t)\hat{s}_{-\mathbf{q}-\mathbf{Q}_{s}}^{-}(0)\right \rangle ,
\]%
where $\hat{s}_{\mathbf{q}}^{\pm }$ is the sum over all the
five-orbital
\begin{figure}[tbp]
\includegraphics[clip,width=0.6\textwidth]{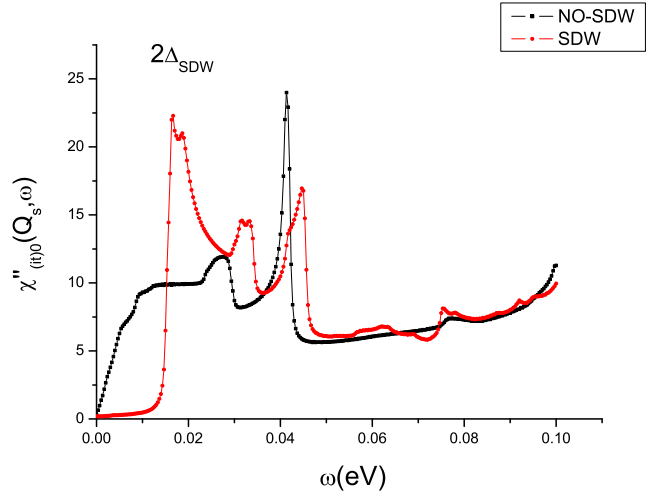}
\caption{(Color online) Dynamic spin susceptibility of itinerant electrons
at the mean-field level, i.e. $\protect\chi _{\left( \mathrm{it}\right) 0%
\mathbf{Q}_{s}}^{\prime \prime }\left( \protect\omega \right) \equiv -2%
\mathrm{Im}\protect\chi _{\left( \mathrm{it}\right) 0}^{+-}\left( \mathbf{Q}%
_{s},\mathbf{Q}_{s},\protect\omega \right) $ with and without SDW order. A
gap $2\Delta _{\mathrm{SDW}}$ in the SDW ordered case is clearly shown.}
\label{Fig4}
\end{figure}
\begin{equation}
\hat{s}_{\mathbf{q}}^{\pm }=\sum_{m}\hat{s}_{\mathbf{q}m}^{\pm }.  \label{sp}
\end{equation}%
In the above mean-field state, after a straightforward but tedious
calculation one obtains, for example,%
\begin{eqnarray}
\chi _{\left( \mathrm{it}\right) 0}^{+-}\left( \mathbf{q},\mathbf{q,}\omega
\right) &=&\sum_{\mathbf{k}\in R,\mathbf{k}+\mathbf{q}\in R,\text{ }E_{%
\mathbf{k},\alpha }>E_{F},\text{ }E_{\mathbf{k}+\mathbf{q},\beta }<E_{F}}%
\frac{\left \vert V_{\mathbf{k},\mathbf{k}+\mathbf{q}}^{\left( 1\right)
}\left( \alpha ,\beta \right) \right \vert ^{2}}{\omega -\left( E_{\mathbf{k}%
,\alpha }-E_{\mathbf{k}+\mathbf{q},\beta }\right) +i0^{+}}  \nonumber \\
&&+\sum_{\mathbf{k}\in R,\mathbf{k}+\mathbf{q}\notin R,\text{ }E_{\mathbf{k}%
,\alpha }>E_{F},\text{ }E_{\mathbf{k}+\mathbf{q\pm Q_{s}},\beta }<E_{F}}%
\frac{\left \vert V_{\mathbf{k},\mathbf{k}+\mathbf{q}\pm \mathbf{Q}%
_{s}}^{\left( 2\right) }\left( \alpha ,\beta \right) \right \vert ^{2}}{%
\omega -\left( E_{\mathbf{k},\alpha }-E_{\mathbf{k}+\mathbf{q\pm Q_{s}}%
,\beta }\right) +i0^{+}}  \nonumber \\
&&+\sum_{\mathbf{k}\in R,\mathbf{k}-\mathbf{q}\in R,\text{ }E_{\mathbf{k}%
,\alpha }>E_{F},\text{ }E_{\mathbf{k}-\mathbf{q},\beta }<E_{F}}\frac{\left
\vert V_{\mathbf{k}-\mathbf{q},\mathbf{k}}^{\left( 1\right) }\left( \beta
,\alpha \right) \right \vert ^{2}}{-\omega -\left( E_{\mathbf{k},\alpha }-E_{%
\mathbf{k}-\mathbf{q},\beta }\right) +i0^{+}}  \nonumber \\
&&+\sum_{\mathbf{k}\in R,\mathbf{k}-\mathbf{q}\notin R,\text{ }E_{\mathbf{k}%
,\alpha }>E_{F},\text{ }E_{\mathbf{k}-\mathbf{q\pm Q_{s}},\beta }<E_{F}}%
\frac{\left \vert V_{\mathbf{k}-\mathbf{q}\pm \mathbf{Q}_{s},\mathbf{k}%
}^{\left( 2\right) }\left( \beta ,\alpha \right) \right \vert ^{2}}{-\omega
-\left( E_{\mathbf{k},\alpha }-E_{\mathbf{k}-\mathbf{q\pm Q_{s}},\beta
}\right) +i0^{+}},  \label{GRFIT2}
\end{eqnarray}%
for the momentum $\mathbf{q}$ within the reduced BZ, i.e. $\mathbf{q}\in R$,
and
\begin{eqnarray}
V_{\mathbf{k},\mathbf{q}}^{\left( 1\right) }\left( \alpha ,\beta \right)
&\equiv &\sum_{m}\left[ U_{\mathbf{k}\downarrow }^{\ast }\left( m,\alpha
\right) U_{\mathbf{q}\uparrow }\left( m,\beta \right) +U_{\mathbf{k}%
\downarrow }^{\ast }\left( m+5,\alpha \right) U_{\mathbf{q}\uparrow }\left(
m+5,\beta \right) \right]  \nonumber \\
V_{\mathbf{k},\mathbf{q}}^{\left( 2\right) }\left( \alpha ,\beta \right)
&\equiv &\sum_{m}\left[ U_{\mathbf{k}\downarrow }^{\ast }\left( m,\alpha
\right) U_{\mathbf{q}\uparrow }\left( m+5,\beta \right) +U_{\mathbf{k}%
\downarrow }^{\ast }\left( m+5,\alpha \right) U_{\mathbf{q}\uparrow }\left(
m,\beta \right) \right] ,  \label{v1v2}
\end{eqnarray}%
with, say,
\begin{equation}
\hat{s}_{-\mathbf{q}}^{-}=\sum_{\alpha \beta ,\mathbf{k}\in R,\mathbf{k}+%
\mathbf{q}\in R}^{{}}c_{\mathbf{k},\alpha ,\downarrow }^{\dagger }c_{\mathbf{%
k}+\mathbf{q},\beta \uparrow }V_{\mathbf{k},\mathbf{k}+\mathbf{q}}^{\left(
1\right) }\left( \alpha ,\beta \right) +\sum_{\alpha \beta ,\mathbf{k}\in R,%
\mathbf{k}+\mathbf{q}\notin R}^{{}}c_{\mathbf{k},\alpha ,\downarrow
}^{\dagger }c_{\mathbf{k}+\mathbf{q}\pm \mathbf{Q}_{s},\beta \uparrow }V_{%
\mathbf{k},\mathbf{k}+\mathbf{q}\pm \mathbf{Q}_{s}}^{\left( 2\right) }\left(
\alpha ,\beta \right) .
\end{equation}%
The other Green's function can be similarly obtained. Numerical calculation
for the dynamic spin susceptibility of the itinerant-electrons at the
mean-field level, i.e. $\chi _{\left( \mathrm{it}\right) 0\mathbf{Q}%
_{s}}^{\prime \prime }\left( \omega \right) \equiv -2\mathrm{Im}\chi
_{\left( \mathrm{it}\right) 0}^{+-}\left( \mathbf{Q}_{s},\mathbf{Q}%
_{s},\omega \right) $ is presented in Fig. 4 with and without SDW order,
where the suppressing of the low-frequency spectrum by the SDW gap at $%
2\Delta _{\mathrm{SDW}}$ is clearly shown.

\section{Spin Dynamics}

In this section, we shall calculate the dynamic spin susceptibility at the
RPA level beyond the above mean-field approximation, by which both a
Goldstone mode and a gapped collective spin mode in the AF ordered phase can
be recovered.

\subsection{RPA treatment}

\begin{figure}[tbp]
\includegraphics[clip,width=0.6\textwidth]{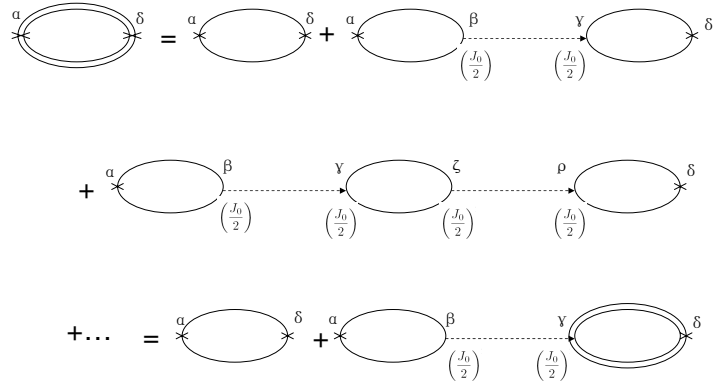}
\caption{(Color online) Feynman diagrams of RPA for dynamic spin
correlation of itinerant electrons. Here the single line bubbles
denote the dynamic spin
correlation function of (free) itinerant electrons, $\hat{\protect\chi}%
_{\left( \mathrm{it}\right) 0}^{+-}\left(
\mathbf{q},\protect\omega \right) , $ the double line bubbles are
the renormalized dynamic spin correlation function of itinerant
electrons, $\hat{\protect\chi}_{\left( \mathrm{it}\right)
}^{+-}\left( \mathbf{q},\protect\omega \right) ,$ the dotted lines
are the dynamic spin correlation function of local moments,
$\hat{\protect\chi}_{\left( \mathrm{lo}\right) 0}^{+-}\left(
\mathbf{q},\protect\omega \right) $. The indices $\protect\alpha
,$ $\protect\beta $, $\cdots\cdots$ label the matrix elements
defined in (\protect\ref{matrixgrf0}).} \label{Fig5}
\end{figure}

The Hund's rule interaction between the local moments and itinerant
electrons in (\ref{hjh}) is rewritten as
\begin{equation}
H_{J_{H}}=-J_{0}\sum \nolimits_{\mathbf{q}}\mathbf{\hat{S}}_{\mathbf{q}%
}\cdot \mathbf{\hat{s}}_{-\mathbf{q}}  \label{rpain}
\end{equation}%
where the spin operator $\mathbf{\hat{s}}$ appears as a whole for the
five-band. At the RPA level, for example,
\begin{eqnarray}
\chi _{\left( \mathrm{it}\right) }^{+-}(\mathbf{q},\mathbf{q,}t) &=&-i\left
\langle T\hat{s}_{\mathbf{q}}^{+}\left( t\right) \hat{s}_{-\mathbf{q}%
}^{-}\right \rangle _{0}+\frac{i}{2}\left \langle T\hat{s}_{\mathbf{q}%
}^{+}\left( t\right) H_{J_{H}}\left( t_{1}\right) H_{J_{H}}\left(
t_{2}\right) \hat{s}_{-\mathbf{q}}^{-}\right \rangle _{0}+O\left(
H_{J_{H}}^{4}\right)  \nonumber \\
&=&\chi _{\left( \mathrm{it}\right) 0}^{+-}\left( \mathbf{q},\mathbf{q,}%
t\right) +\left( \frac{J_{0}}{2}\right) ^{2}\int_{-\infty }^{\infty
}dt_{1}\int_{-\infty }^{\infty }dt_{2}[\chi _{\left( \mathrm{it}\right)
0}^{+-}\left( \mathbf{q},\mathbf{q,}t-t_{1}\right) \chi _{\left( \mathrm{lo}%
\right) 0}^{+-}\left( \mathbf{q},\mathbf{q,}t_{1}-t_{2}\right) \chi _{\left(
\mathrm{it}\right) 0}^{+-}\left( \mathbf{q},\mathbf{q,}t_{2}\right)
\nonumber \\
&&+\chi _{\left( \mathrm{it}\right) 0}^{+-}\left( \mathbf{q},\mathbf{q,}%
t-t_{1}\right) \chi _{\left( \mathrm{lo}\right) 0}^{+-}\left( \mathbf{q},%
\mathbf{q,}t_{1}-t_{2}\right) \chi _{\left( \mathrm{it}\right) 0}^{+-}\left(
\mathbf{q+\mathbf{Q}_{s}},\mathbf{q,}t_{2}\right)  \nonumber \\
&&+\chi _{\left( \mathrm{it}\right) 0}^{+-}\left( \mathbf{q},\mathbf{q+%
\mathbf{Q}_{s},}t-t_{1}\right) \chi _{\left( \mathrm{lo}\right)
0}^{+-}\left( \mathbf{q+\mathbf{Q}_{s}},\mathbf{q+\mathbf{Q}_{s},}%
t_{1}-t_{2}\right) \chi _{\left( \mathrm{it}\right) 0}^{+-}\left( \mathbf{q+%
\mathbf{Q}_{s}},\mathbf{q,}t_{2}\right)  \nonumber \\
&&+\chi _{\left( \mathrm{it}\right) 0}^{+-}\left( \mathbf{q},\mathbf{q+%
\mathbf{Q}_{s},}t-t_{1}\right) \chi _{\left( \mathrm{lo}\right)
0}^{+-}\left( \mathbf{q+\mathbf{Q}_{s}},\mathbf{q,}t_{1}-t_{2}\right) \chi
_{\left( \mathrm{it}\right) 0}^{+-}\left( \mathbf{q},\mathbf{q,}t_{2}\right)
]+...,  \label{matrixgrf}
\end{eqnarray}%
which is illustrated diagrammatically by Fig. \ref{Fig5}.

To make the formulation more compact, we define a $2\times 2$ matrix $\hat{%
\chi}^{+-}\left( \mathbf{q,}\omega \right) $ with the components
\begin{eqnarray}
\hat{\chi}_{\left( 1,1\right) }^{+-}\left( \mathbf{q,}\omega \right) &\equiv
&\chi ^{+-}\left( \mathbf{q},\mathbf{q,}\omega \right)  \nonumber \\
\hat{\chi}_{\left( 1,2\right) }^{+-}\left( \mathbf{q,}\omega \right) &\equiv
&\chi _{{}}^{+-}\left( \mathbf{q},\mathbf{q}+\mathbf{Q}_{s},\omega \right)
\nonumber \\
\hat{\chi}_{\left( 2,2\right) }^{+-}\left( \mathbf{q,}\omega \right) &\equiv
&\chi _{{}}^{+-}\left( \mathbf{q}+\mathbf{Q}_{s},\mathbf{q}+\mathbf{Q}%
_{s},\omega \right)  \nonumber \\
\hat{\chi}_{\left( 2,1\right) }^{+-}\left( \mathbf{q,}\omega \right) &\equiv
&\chi _{{}}^{+-}\left( \mathbf{q}+\mathbf{Q}_{s},\mathbf{q,}\omega \right) .
\label{matrixgrf0}
\end{eqnarray}%
With this definition, the Fourier transformation of (\ref{matrixgrf}) reads
\begin{equation}
\hat{\chi}_{\left( \mathrm{it}\right) \left( 1,1\right) }^{+-}=\hat{\chi}%
_{\left( \mathrm{it}\right) 0\left( 1,1\right) }^{+-}+\left( \frac{J_{0}}{2}%
\right) ^{2}\sum_{i,j=1,2}\hat{\chi}_{\left( \mathrm{it}\right)
0\left( 1,i\right) }^{+-}\hat{\chi}_{\left( \mathrm{lo}\right)
0\left( i,j\right) }^{+-}\hat{\chi}_{\left( \mathrm{it}\right)
0\left( j,1\right) }^{+-}+\cdots ,
\end{equation}%
and more generally
\begin{equation}
\hat{\chi}_{\left( \mathrm{it}\right) \left( i,j\right) }^{+-}=\hat{\chi}%
_{\left( \mathrm{it}\right) 0\left( i,j\right) }^{+-}+\left( \frac{J_{0}}{2}%
\right) ^{2}\sum_{i^{\prime },j^{\prime }=1,2}\hat{\chi}_{\left( \mathrm{it}%
\right) 0\left( i,i^{\prime }\right) }^{+-}\hat{\chi}_{\left( \mathrm{lo}%
\right) 0\left( i^{\prime },j^{\prime }\right) }^{+-}\hat{\chi}_{\left(
\mathrm{it}\right) 0\left( j^{\prime },j\right) }^{+-}+\cdots .
\end{equation}

\begin{figure}[tbp]
\includegraphics[clip,width=0.6\textwidth]{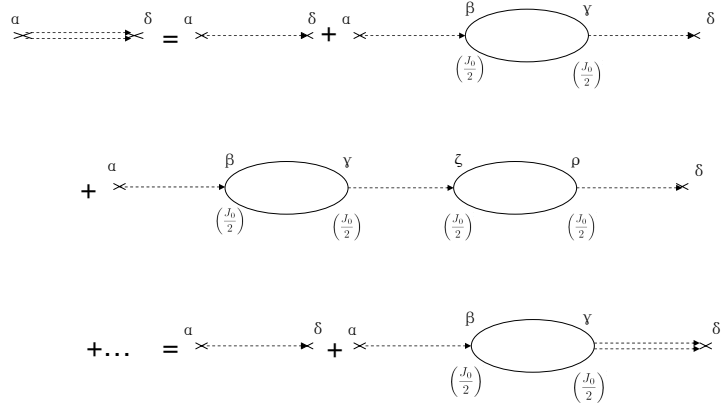}
\caption{(Color online) Feynman diagrams of RPA for dynamic spin
correlation of local moments. Here the double dotted lines are the
renormalized dynamic spin correlation function of local-moments
$\hat{\protect\chi}_{\left( \mathrm{lo}\right) }^{+-}\left(
\mathbf{q},\protect\omega \right) $. The other symbols are the
same as in Fig. 5.} \label{Fig6}
\end{figure}

As the Dyson equations in a compact matrix form, the above RPA result for
the itinerant electrons can be reexpressed as
\begin{equation}
\hat{\chi}_{\left( \mathrm{it}\right) }^{+-}\left( \mathbf{q},\omega \right)
=\left[ I-(\frac{J_{0}}{2})^{2}\hat{\chi}_{\left( \mathrm{it}\right)
0}^{+-}\left( \mathbf{q},\omega \right) \hat{\chi}_{\left( \mathrm{lo}%
\right) 0}^{+-}\left( \mathbf{q},\omega \right) \right] ^{-1}\hat{\chi}%
_{\left( \mathrm{it}\right) 0}^{+-}\left( \mathbf{q},\omega \right) ,
\label{solvedysonit}
\end{equation}%
and similarly for the local moment part illustrated by Fig. \ref{Fig6}, we
have
\begin{equation}
\hat{\chi}_{\left( \mathrm{lo}\right) }^{+-}\left( \mathbf{q},\omega \right)
=\left[ I-(\frac{J_{0}}{2})^{2}\hat{\chi}_{\left( \mathrm{lo}\right)
0}^{+-}\left( \mathbf{q},\omega \right) \hat{\chi}_{\left( \mathrm{it}%
\right) 0}^{+-}\left( \mathbf{q},\omega \right) \right] ^{-1}\hat{\chi}%
_{\left( \mathrm{lo}\right) 0}^{+-}\left( \mathbf{q},\omega \right) ,
\label{solvedysonlo}
\end{equation}%
with the total dynamic spin susceptibility matrix given by
\begin{equation}
\hat{\chi}_{\mathrm{RPA}}^{+-}\left( \mathbf{q},\omega \right) =\hat{\chi}%
_{\left( \mathrm{it}\right) }^{+-}\left( \mathbf{q},\omega \right) +\hat{\chi%
}_{\left( \mathrm{lo}\right) }^{+-}\left( \mathbf{q},\omega \right) .
\label{solvedysontot}
\end{equation}

\begin{figure}[tbp]
\includegraphics[clip,width=0.6\textwidth]{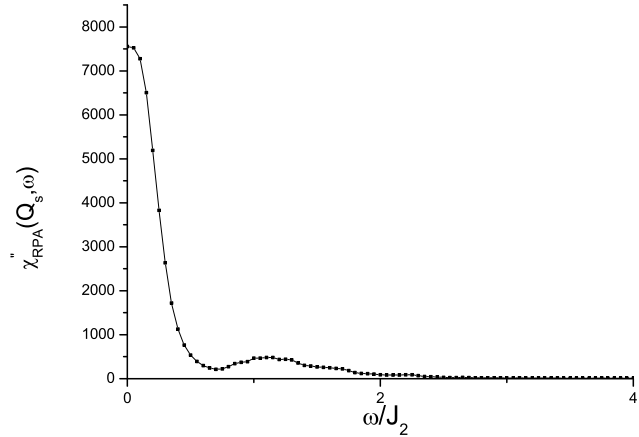}
\caption{(Color online) Total dynamic spin-susceptibility of coupled system
in RPA level $\protect\chi _{\mathrm{RPA}}^{\prime \prime }(\mathbf{q},%
\protect\omega )$ fixing $\mathbf{q}=\mathbf{Q}_{s}$.}
\label{Fig7}
\end{figure}

\subsection{Collective spin modes}

Now we focus on the total dynamic spin susceptibility defined by
\begin{equation}
\chi _{\mathrm{RPA}}^{\prime \prime }\left( \mathbf{q},\omega \right) =-2%
\mathrm{Im}\hat \chi _{\mathrm{RPA}\left(1,1\right)}^{+-}\left( \mathbf{q}%
,\omega \right)  \label{spinexitationspectrum}
\end{equation}%
which can be numerically determined based on the RPA expressions (\ref%
{solvedysonit}), (\ref{solvedysonlo}) and (\ref{solvedysontot}) given in the
above subsection.

In Fig. \ref{Fig7}, $\chi _{\mathrm{RPA}}^{\prime \prime }(\mathbf{q},\omega
)$ at a fixing $\mathbf{q}=\mathbf{Q}_{s}$ is shown as a function of $\omega
$. Here we find a sharp peak emerges at $\omega =0$, in contrast to Figs. %
\ref{Fig3} and \ref{Fig4}. Namely a zero-mode (Goldstone mode) pole is
indeed restored at $\mathbf{q}=\mathbf{Q}_{s}$ in the RPA spin-spin
correlation function. Mathematically, it originates from the vanishing
denominator in (\ref{solvedysonit}) and (\ref{solvedysonlo}).

It is interesting to note that besides the Goldstone mode, there remains a
high-energy mode as represented by the hump in Fig. \ref{Fig7}. It can be
traced to the pole of $\hat{\chi}_{\left( \mathrm{lo}\right) 0}^{+-}\left(
\mathbf{q},\omega \right) $ (cf. Fig. \ref{Fig3}), only broadened through
the scattering with the itinerant electrons. In other words, the gapped
collective mode of the local moments identified in the mean-field state in
the previous section is still present at the RPA level. Physically this
gapped mode is an \textquotedblleft out of phase\textquotedblright \
fluctuations of the local moments relative to the magnetization of itinerant
electrons, while the Goldstone mode is the \textquotedblleft in
phase\textquotedblright \ fluctuations of the locked magnetizations from the
local moments and itinerant electrons.

To display the spin dynamics of the system in the whole BZ, the calculated $%
\chi _{\mathrm{RPA}}^{\prime \prime }\left( \mathbf{q},\omega \right) $ is
presented in Fig. 8(a) with the $x$-axis representing the momentum $\mathbf{q%
}$ along the high-symmetry lines in the reduced BZ and the $y$-axis for the
frequency $\omega$, while the brightness depicting the spectral weight $\chi
_{\mathrm{RPA}}^{\prime \prime }\left( \mathbf{q},\omega \right) $. From
Fig. 8(a), it is clearly shown that the spin excitation spectrum in the AF
ordered phase is separated into two branches, i.e. the lower Goldstone-mode
branch and the upper \textquotedblleft out-of-phase\textquotedblright \ mode
branch. The dispersions of the two modes are illustrated in Fig. 8(b), which
are defined by $\left( \mathbf{q},\omega \right) $ at the largest
(brightest) $\chi _{\mathrm{RPA}}^{\prime \prime }\left( \mathbf{q},\omega
\right) $ in Fig. 8(a). The spin wave dispersion of the pure $J_{1}-J_{2}$
model, i.e., (\ref{bgtr3}) and that of the gapped one, (\ref%
{dispersionspinwavemeanfield}), due to coupling to the SDW order of the
itinerant electrons are also shown for comparison.

\begin{figure}[tbp]
\includegraphics[clip,width=0.8\textwidth]{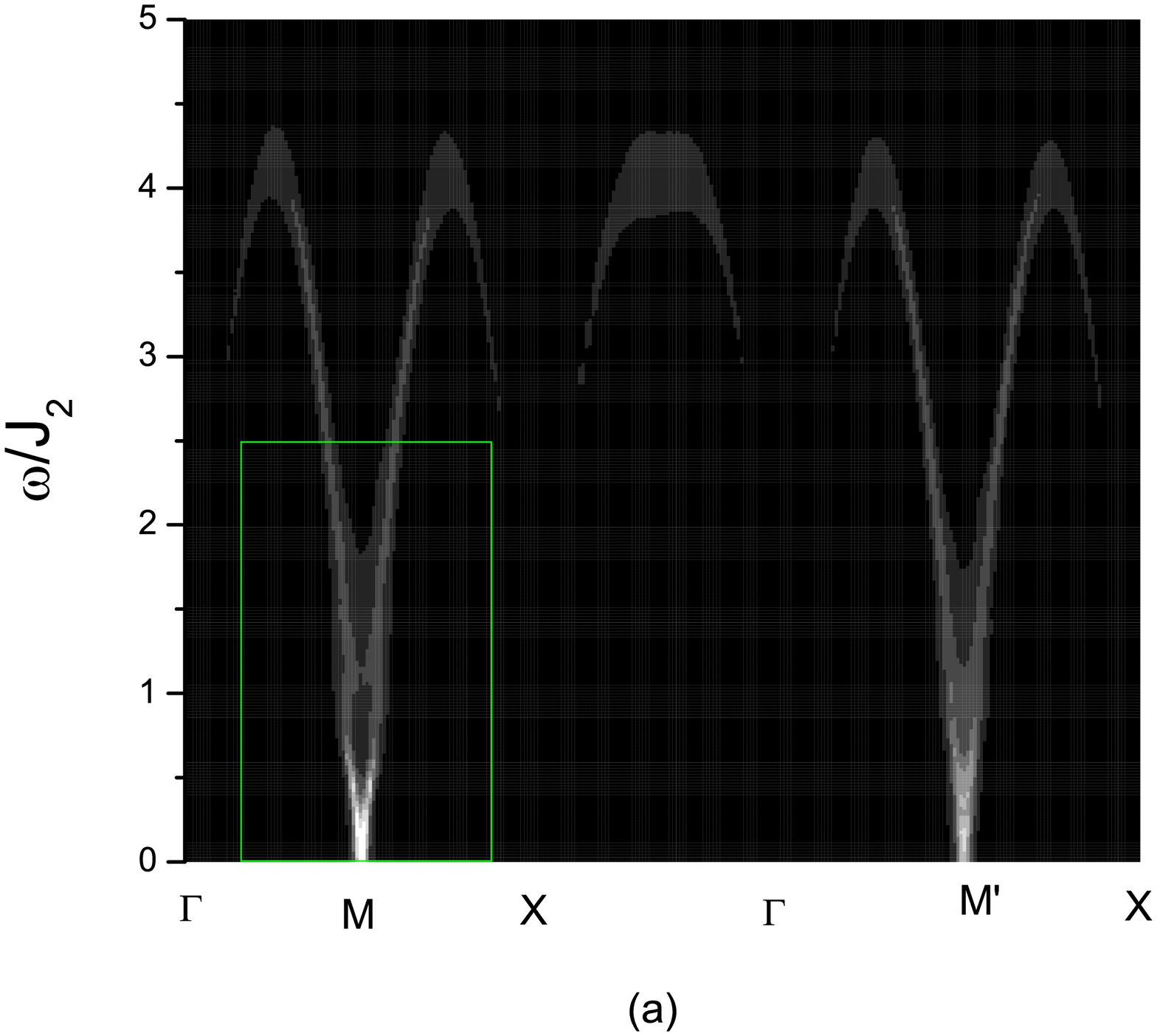} %
\includegraphics[clip,width=0.6\textwidth]{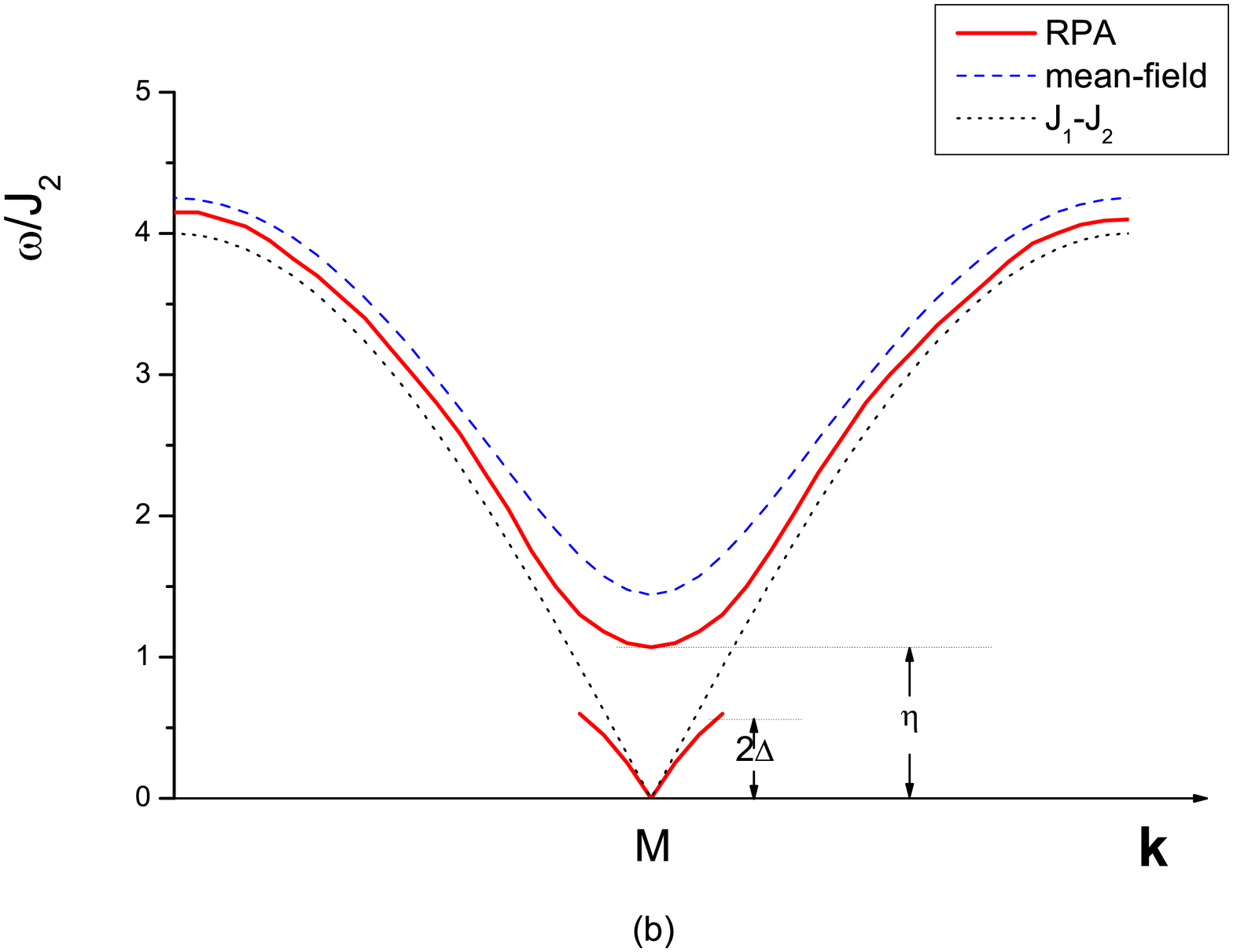}
\caption{(Color online) (a) The brightness represents the spectral weight of
the dynamic spin susceptibility $\protect\chi _{\mathrm{RPA}}^{\prime \prime
}$ in the $\mathbf{q}$ and $\protect\omega $ space. (b) The dispersions of
two branches of the collective spin mode as read from (a). The dispersions
of the spin-wave of the pure $J_{1}-J_{2}$ model (dotted) and that with a
gap opening at the mean-field level (dashed) are also shown for comparison. }
\label{Fig.8}
\end{figure}

The lower branch Goldstone mode in Fig. 8(b) is well-defined within a small
region which centers at the AF wavevector $\mathbf{Q}_{s}$, upper-bounded by
the SDW gap, i.e. $2\Delta _{\mathrm{SDW}}$ of the itinerant electrons,
beyond which it decays quickly due to the scattering with the particle-hole
continuum of itinerant electrons (cf. Fig. \ref{Fig4}). Inside the SDW order
gap, such a Goldstone mode is protected and is decoupled from the itinerant
electrons due to the reconstruction of the Fermi surface shown in Figs. \ref%
{Fig.1} and \ref{Fig.2}.

On the other hand, the high energy \textquotedblleft
out-of-phase\textquotedblright \ mode is mainly contributed by the local
moments and present throughout the BZ. It gets slight renormalization and
broadening due to the scattering from the itinerant electrons at the RPA
level, but more or less follows the dispersion obtained in the mean-field
treatment (dotted), with a gap $\eta $
\begin{equation}
\eta \approx \omega _{\mathbf{Q}_{s}}.  \label{gap}
\end{equation}

\begin{figure}[tbp]
\includegraphics[clip,width=0.8\textwidth]{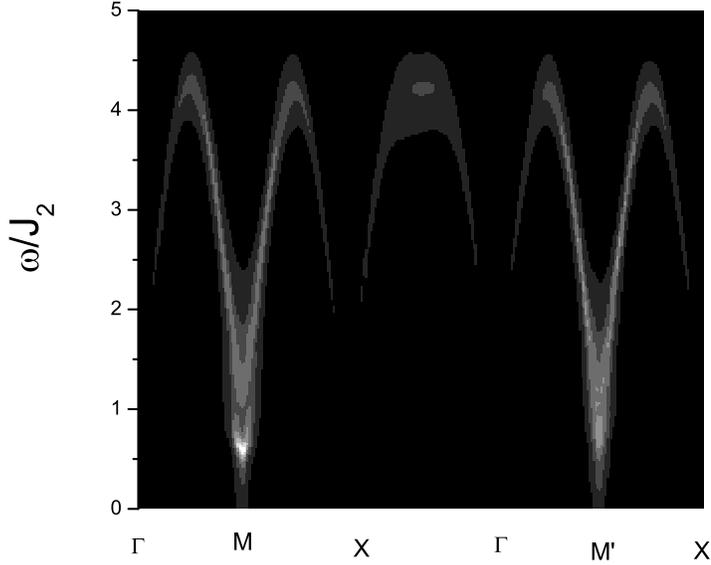}
\caption{(Color online) The dynamic spin susceptibility similar to Fig.
\protect\ref{Fig.8}, but with an additional ion-anisotropy $J_{z}=1$ \textrm{%
meV}.}
\label{Fig.9}
\end{figure}

It is noted that the Goldstone mode will be fragile against the
ion-anisotropy. By adding an ion-anisotropy term,
\begin{equation}
H_{\left( \mathrm{ion}\right) }=-J_{z}\sum_{i}\left( S_{i}^{z}\right) ^{2}
\label{Jz}
\end{equation}%
the spin-rotational symmetry will be broken in $H_{\mathrm{lo}}$. By using
the spin-wave expansion
\begin{equation}
H_{\left( \mathrm{ion}\right) }=2SJ_{z}\sum \nolimits_{k}^{\prime }\left( a_{%
\mathbf{k}}^{\dagger }a_{\mathbf{k}}+b_{\mathbf{k}}^{\dagger }b_{\mathbf{k}%
}\right) +\mathrm{const.}  \label{Jzsp}
\end{equation}%
one finds the spin-excitation spectrum (\ref{bgtr3}) acquires an
ion-anisotropy gap
\begin{equation}
\omega _{\mathbf{k}}=S\sqrt{\left( \Gamma _{\mathbf{k}}+2J_{z}\right)
^{2}-M_{\mathbf{k}}^{2}}.  \label{disJz}
\end{equation}%
At the RPA level, the Goldstone spin wave will also become gapped
and its spectrum weight is reduced with reference to that without
ion-anisotropy, as clearly shown in Fig. \ref{Fig.9}. When one
further increases $J_{z},$ the Goldstone mode disappears, but the
gapped \textquotedblleft out-of-phase\textquotedblright \
collective mode remains robust and insensitive to the
ion-anisotropy.

Therefore, in the present coupled local moment and itinerant electron
system, the Goldstone theorem still holds for a spontaneously symmetry
breaking state with AF ordering at the RPA level. However, the Goldstone
mode is very sensitive to the presence of ion-anisotropy and does not play
an important role for charge dynamics as it is decoupled from the itinerant
electrons. On the other hand, the gapped \textquotedblleft
out-of-phase\textquotedblright \ collective mode is more prominent which
extends over the whole BZ with a spin-wave bandwidth $\sim $ $4J_{2}$ and
can be easily probed by neutron scattering experiments. Such two-branch
collective spin excitations are unique prediction for the AF ordered phase.
In the next section, the charge response will be further examined based on
the scattering of such collective spin modes with the itinerant electrons in
SDW ordering.

\section{Charge dynamics}

In this section, we study the charge dynamics in the coupled local-moment
and itinerant-electron system in the AF ordered phase. Here the itinerant
electrons in the Fermi surface region will be highly coherent even in the
presence of the gapless Goldstone mode.

\subsection{Self-energy of itinerant electrons}

By going beyond the mean-field linearization in (\ref{couple}), we shall
consider the scattering based on
\begin{eqnarray}
H_{J_{H}} &\rightarrow &H_{I}^{\prime }=-\frac{J_{0}}{2\sqrt{N}}\sum_{%
\mathbf{k}\in R,\mathbf{q}\in R,\alpha \beta }c_{\mathbf{k}\alpha \downarrow
}^{\dagger }c_{\mathbf{q}\beta \uparrow }\left[ S_{\mathbf{q}-\mathbf{k}%
}^{+}V_{\mathbf{k},\mathbf{q}}^{\left( 1\right)}\left(\alpha,\beta
\right)+S_{\mathbf{q}-\mathbf{k}+\mathbf{Q}_{s}}^{+}V_{\mathbf{k},\mathbf{q}%
}^{\left( 2\right) }\left(\alpha,\beta \right)\right] +h.c.  \nonumber \\
&\equiv &-\frac{J_{0}}{2\sqrt{N}}\sum_{\mathbf{k}\in R,\mathbf{q}\in
R,\alpha \beta }c_{\mathbf{k}\alpha \downarrow }^{\dagger }c_{\mathbf{q}%
\beta \uparrow }R_{\mathbf{kq}\alpha \beta }^{+}+h.c.  \label{scatter}
\end{eqnarray}%
where the $V_{\mathbf{k},\mathbf{q}}^{\left(1, 2\right) }\left(\alpha,\beta
\right)$ are defined by (\ref{v1v2}). Here we retain only the scattering of
itinerant electrons with the transverse fluctuations of local moments as the
longitudinal fluctuations in $S^{z}$ are gapped with $\left \langle
S^{z}\right
\rangle \neq 0$.

The single-particle Green's Function can be evaluated perturbatively by
\begin{equation}
G_{\alpha \beta \downarrow }(\mathbf{k},\omega )=G_{\alpha \beta \downarrow
}^{0}\left( \mathbf{k},\omega \right) +\sum_{\gamma \epsilon }G_{\alpha
\gamma \downarrow }^{0}(\mathbf{k},\omega )\Sigma _{\gamma \epsilon
\downarrow }(\mathbf{k},\omega )G_{\epsilon \beta \downarrow }^{0}(\mathbf{k}%
,\omega )+O\left( H_{I}^{\prime 4}\right) ,  \label{spgrfnw}
\end{equation}%
where the self-energy%
\begin{equation}
\Sigma _{\gamma \epsilon \downarrow }(\mathbf{k},\omega )=\frac{iJ_{0}^{2}}{%
4N}\sum_{\mathbf{q}\in R,\theta \xi }\int_{-\infty }^{+\infty }\frac{d\Omega
}{2\pi }G_{\theta \xi \uparrow }^{0}\left( \mathbf{q},\omega -\Omega \right)
\chi _{\left( R\right) \mathbf{kq}\gamma \theta \xi \epsilon }^{+-}\left(
\Omega \right) ,  \label{selfenergyw}
\end{equation}%
with
\begin{eqnarray}
\chi _{\left( R\right) \mathbf{kq}\gamma \theta \xi \epsilon }^{+-}\left(
\Omega \right) &=&\hat{\chi}_{\left( \mathrm{lo}\right) \left( 1,1\right)
}^{+-}\left( \mathbf{q}-\mathbf{k},\Omega \right) \left \vert V^{\left(
1\right) }\right \vert ^{2}+\hat{\chi}_{\left( \mathrm{lo}\right) \left(
1,2\right) }^{+-}\left( \mathbf{q}-\mathbf{k},\Omega \right) V^{\left(
1\right) }V^{\left( 2\right) \ast }+  \nonumber \\
&&\hat{\chi}_{\left( \mathrm{lo}\right) \left( 2,1\right) }^{+-}\left(
\mathbf{q}-\mathbf{k},\Omega \right) V^{\left( 2\right) }V^{\left( 1\right)
\ast }+\hat{\chi}_{\left( \mathrm{lo}\right) \left( 2,2\right) }^{+-}\left(
\mathbf{q}-\mathbf{k},\Omega \right) \left \vert V^{\left( 2\right) }\right
\vert ^{2}.  \label{sppropogaterw}
\end{eqnarray}%
Here the matrix $\hat{\chi}_{\left( \mathrm{lo}\right) }^{+-}$ is
the dynamic spin correlation function for the local-moment in the RPA level defined in (\ref%
{solvedysonlo}). Fig. 10 and Fig. 11 show the corresponding
Feynman diagrams of the self-energy correction for itinerant
electrons.

\begin{figure}[tbp]
\includegraphics[clip,width=0.6\textwidth]{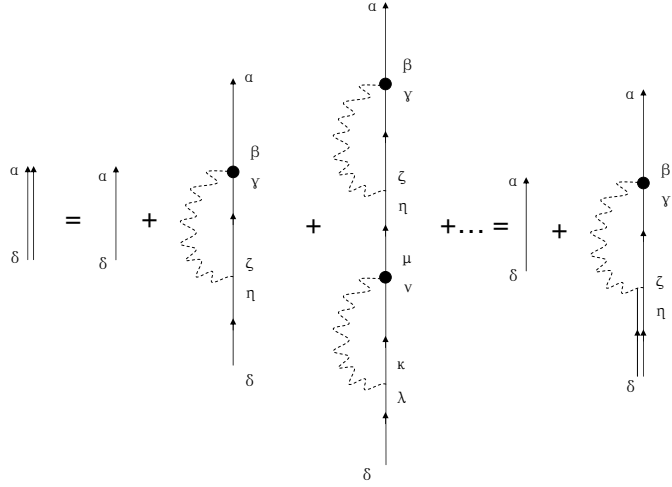}
\caption{(Color online) Feynman diagrams of self-energy correction
for itinerant-electrons. Here the single black lines denote the
proporgators of (free)
itinerant-electrons, $G_{\protect\alpha \protect\delta }^{0}(\mathbf{k},%
\protect\omega ),$ \ the double black lines are the renormalized
proporgators of itinerant-electrons, $G_{\protect\alpha \protect\delta }(%
\mathbf{k},\protect\omega )$, the dotted wavy lines are the
modified dynamic spin correlation of local-moments in the RPA level $\protect\chi %
_{\left( R\right) \mathbf{kq}\protect\gamma \protect\theta \protect\xi
\protect\epsilon }^{+-}\left( \Omega \right) $ defined by (\protect\ref%
{sppropogaterw}). The indices $\protect\alpha , $ $\protect\beta ,$ $\protect%
\gamma ,$ $\protect\rho ,$ $\protect\varsigma ,$ label different bands. }
\label{Fig.10}
\end{figure}

The equation (\ref{spgrfnw}) can be further expressed in a $10\times 10$
matrix formalism by
\begin{eqnarray}
G_{\mathbf{k},\omega } &=&G_{\mathbf{k},\omega }^{0}+G_{\mathbf{k},\omega
}^{0}\Sigma _{\mathbf{k},\omega }G_{\mathbf{k},\omega }^{0}+\cdots  \nonumber
\\
&=&G_{\mathbf{k},\omega }^{0}+G_{\mathbf{k},\omega }^{0}\Sigma _{\mathbf{k}%
,\omega }G_{\mathbf{k},\omega }  \label{dyson}
\end{eqnarray}%
which gives rise to $G_{\mathbf{k},\omega }=\left( \left( G_{\mathbf{k}%
,\omega }^{0}\right) ^{-1}-\left( \Sigma _{\mathbf{k},\omega }\right)
\right) ^{-1}.$ Here we have omitted the spin index as they are in fact
spin-independent.

\begin{figure}[tbp]
\includegraphics[clip,width=0.6\textwidth]{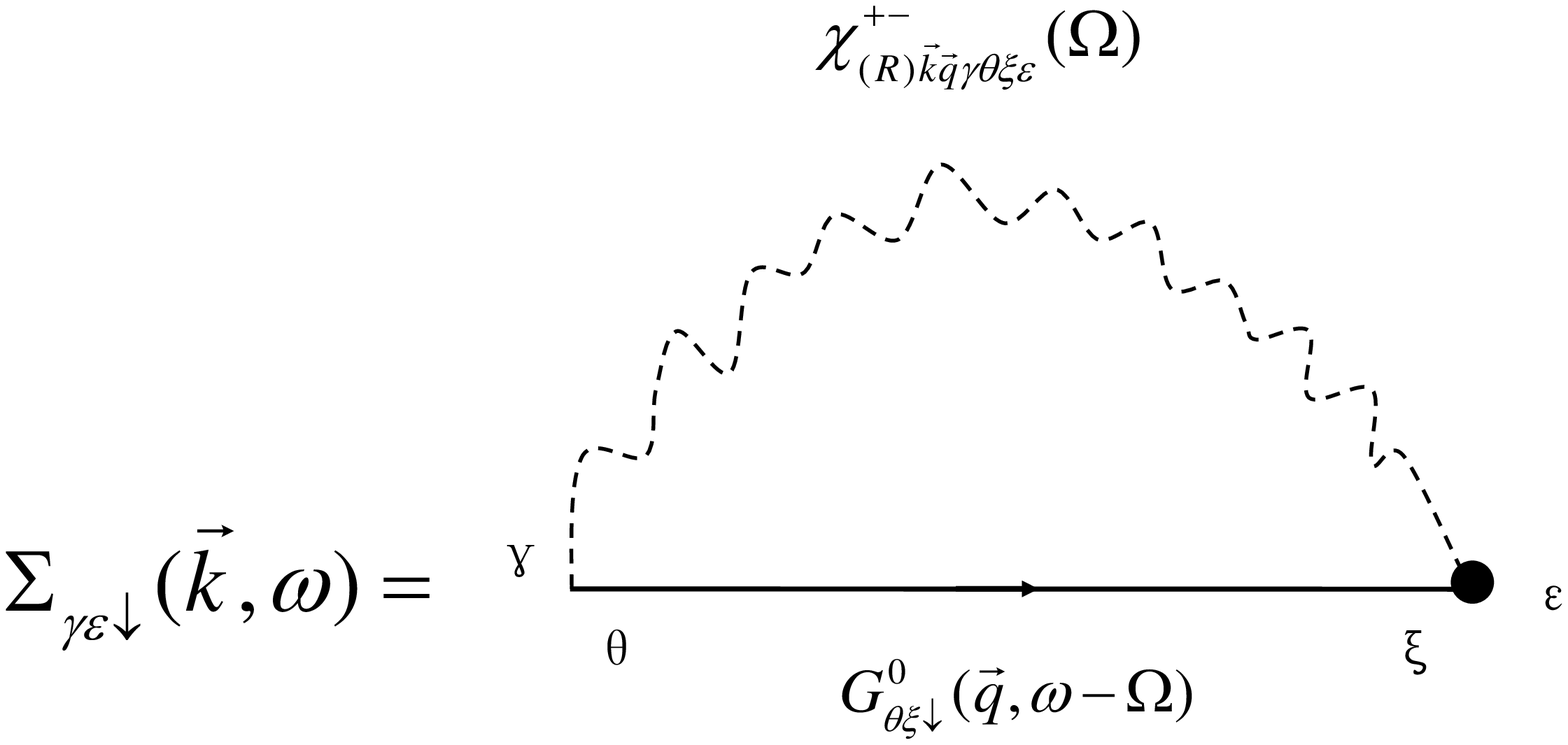}
\caption{(Color online) Self-energy correction for itinerant-electrons }
\label{Fig.11}
\end{figure}
By noting that the zero-th order single-particle Green's function is
diagonal, through the Lehmann representation we have for $\omega >0$
\begin{equation}
-\mathrm{Im}\Sigma _{\gamma \epsilon }(\mathbf{k},\omega )=\frac{J_{0}^{2}}{%
4N}\sum_{\mathbf{q}\in R,\theta }\int_{-\infty }^{+\infty }\frac{d\Omega }{%
4\pi }\rho _{\theta }^{0}\left( \mathbf{q},\omega -\Omega \right) D_{\gamma
\theta \epsilon }^{+-}\left( \mathbf{k}-\mathbf{q},\Omega \right) \left[
n_{B}\left( \Omega \right) +n_{F}\left( \Omega -\omega \right) \right] ,
\label{selfenergy2}
\end{equation}%
where
\begin{equation}
\rho _{\theta }^{0}\left( \mathbf{q},\omega \right) \equiv -2\mathrm{Im}%
G_{\theta \theta }^{0}\left( \mathbf{q},\omega \right)
\label{spectrafunction}
\end{equation}%
and
\begin{equation}
D_{\gamma \theta \epsilon }^{+-}\left( \mathbf{k}-\mathbf{q},\Omega \right)
\equiv -2\mathrm{Im}\chi _{\left( R\right) \mathbf{kq}\gamma \theta \theta
\epsilon }^{+-}\left( \Omega \right) ,  \label{dfunction}
\end{equation}%
are the spectral functions of single-particle and the modified dynamic spin
susceptibility of the local-moment, respectively.

At zero temperature, the imaginary self-energy (\ref{selfenergy2}) can be
further evaluated as (for $\omega >0$)
\begin{equation}
-\mathrm{Im}\Sigma _{\alpha \beta }(\mathbf{k},\omega )=\frac{J_{0}^{2}}{8N}%
\sum_{\mathbf{q}\in R,\theta ,0<E_{\mathbf{q}\theta }<\omega }D_{\alpha
\theta \beta }^{+-}\left( \mathbf{k}-\mathbf{q},\omega -E_{\mathbf{q}\beta
}\right).  \label{selfenwp}
\end{equation}

\begin{figure}[tbp]
\includegraphics[clip,width=0.6\textwidth]{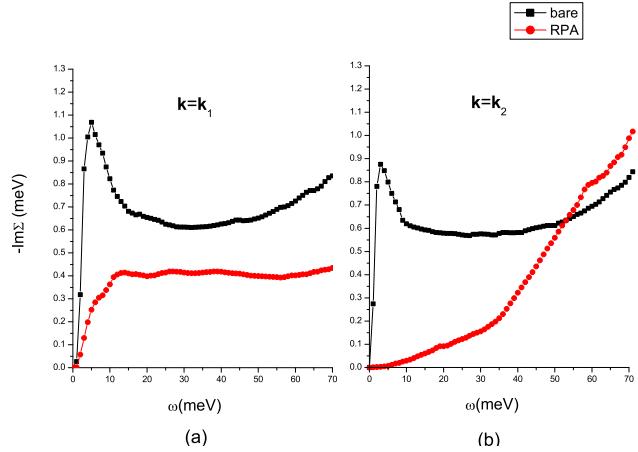}
\caption{(Color online) The imaginary part of the quasiparticle
self-energy -$\mathrm{Im}\Sigma_{\alpha \alpha} (\mathbf{k},\omega
)$ at two typical k-points marked in Fig. 2(a) is shown. The
red-circled curve is for incorporating the scattering with the
full collective mode at the RPA level and the black-squared one is
for the scattering with the bare local moment fluctuations
governed by $H_{\mathrm{lo}}$.} \label{Fig.12}
\end{figure}

In Fig. 12, the -$\mathrm{Im}\Sigma_{\alpha \alpha} (\mathbf{k},\omega )$ as
a function of $\omega $ is shown at two typical momenta, $\left(\mathbf{k}%
_{1},\alpha_{1}\right)$ and $\left(\mathbf{k}_{2},\alpha_{2}\right)$,
respectively, which are marked at Fermi pockets in Fig. 2(a). It shows that
the life time of the quasiparticle excitations at these points of Fermi
pockets actually gets substantially \emph{enhanced} at low $\omega $ in the
collinear AF ordered state (solid circles), as compared to an artificial
case (solid squares) without an induced SDW order appearing in itinerant
electrons such that there is no gap $\eta $ opened up in the
\textquotedblleft out-of-phase\textquotedblright \ mode. It implies that
although in the normal state the itinerant electrons may be strongly
scattered by the low-lying local moment fluctuations, the sharp coherence of
quasiparticles will emerge in the AF state, where the gapless Goldstone mode
is essentially decoupled from the particle-hole continuum and the
\textquotedblleft out-of-phase\textquotedblright \ mode is gapped.

\subsection{Optical conductivity}

Finally, let us examine the overall structure of the optical conductivity $%
\sigma \left( \omega \right) $ at the mean-field and RPA levels. It is
related to the current-current correlation function $G_{J}\left( \omega
\right) $ through the relation
\begin{equation}
\sigma \left( \omega \right) =-\frac{1}{\omega }\mathrm{Im}G_{J}\left(
\omega \right) ,  \label{opticalconductivity}
\end{equation}%
where $G_{J}\left( \omega \right) $ is the Fourier transformation of the
current-current correlator
\begin{equation}
G_{J}\left( t\right) \equiv -i\left \langle TJ\left( t\right) J(0)\right
\rangle .  \label{cccor}
\end{equation}%
Here $J$ denotes the $\mathbf{q}=0$ current operator $\mathbf{J}$ along,
say, the $x$ axis for the five-band model (\ref{hitr}) defined by%
\begin{eqnarray}
\mathbf{J} &=&\frac{\partial H_{it}(\mathbf{A})}{\partial \mathbf{A}}|_{%
\mathbf{A}=0}  \nonumber \\
&=&\sum_{\mathbf{k}\sigma mn}\vec{\bigtriangledown}_{\mathbf{k}}f_{mn}(
\mathbf{k})c_{\mathbf{k}m\sigma }^{+}c_{\mathbf{k}n\sigma }  \nonumber \\
&=&\sum_{\mathbf{k}\in R,\sigma mn \alpha \beta }c_{\mathbf{k}\alpha \sigma
}^{\dagger }c_{\mathbf{k}\beta \sigma }\left[\vec{\bigtriangledown}_{\mathbf{%
k}}f_{mn}( \mathbf{k})U_{\mathbf{k}\sigma}^{\ast }\left( m,\alpha \right) U_{%
\mathbf{k}\sigma}\left(n,\beta \right)+\vec{\bigtriangledown}_{\mathbf{k}%
}f_{mn}(\mathbf{k}+\mathbf{Q}_{s})U_{\mathbf{k}\sigma}^{\ast
}\left(m+5,\alpha \right) U_{\mathbf{k}\sigma}\left(n+5,\beta \right)\right]
\nonumber \\
&\equiv &\sum_{\mathbf{k}\in R,\sigma \alpha \beta }V_{\mathbf{k}\sigma
\alpha \beta }c_{\mathbf{k}\alpha \sigma }^{\dagger }c_{\mathbf{k}\beta
\sigma }.
\end{eqnarray}

\begin{figure}[tbp]
\includegraphics[clip,width=0.6\textwidth]{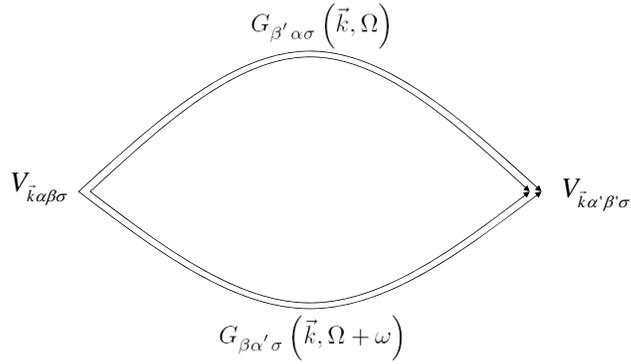}
\caption{(Color online) Current-current correlation function for
itinerant-electrons. Here the double black lines are the
renormalized proporgators of itinerant-electrons. } \label{Fig.13}
\end{figure}

Omitting the vertex-correction, we find
\begin{equation}
G_{J}\left( \omega \right) =-\frac{i}{N}\sum_{\mathbf{k}\in R,\sigma ,\alpha
\beta ,\alpha ^{^{\prime }}\beta ^{^{\prime }}}\int_{-\infty }^{+\infty }%
\frac{d\Omega }{2\pi }V_{\mathbf{k}\sigma \alpha \beta }V_{\mathbf{k}\sigma
\alpha ^{^{\prime }}\beta ^{^{\prime }}}G_{\beta ^{^{\prime }}\alpha \sigma
}\left( \mathbf{k},\Omega \right) G_{\beta \alpha ^{^{\prime }}\sigma
}\left( \mathbf{k},\Omega +\omega \right) .
\end{equation}%
with the Feynman diagrams shown in Fig. 13. Denoting
\begin{equation}
\rho _{\alpha \beta }\left( \mathbf{k},\omega \right) \equiv -2\mathrm{Im}%
G_{\alpha \beta \sigma }\left( \mathbf{k},\omega \right) \mathrm{sgn}\left(
\omega \right) ,  \label{spectrumfunction}
\end{equation}%
at $T=0$ we obtain
\begin{equation}
-\mathrm{Im}G_{J}=\frac{1}{N}\sum_{\mathbf{k}\in R,\sigma ,\alpha \beta
,\alpha ^{^{\prime }}\beta ^{^{\prime }}}\int_{-\omega }^{0}\frac{d\Omega }{%
2\pi }V_{\mathbf{k}\sigma \alpha \beta }V_{\mathbf{k}\sigma \alpha
^{^{\prime }}\beta ^{^{\prime }}}\rho _{\beta ^{^{\prime }}\alpha }\left(
\mathbf{k},\Omega \right) \rho _{\beta \alpha ^{^{\prime }}}\left( \mathbf{k}%
,\Omega +\omega \right) .
\end{equation}%
such that
\begin{equation}
\sigma \left( \omega \right) =\frac{1}{N\omega }\sum_{\mathbf{k}\in R,\sigma
,\alpha \beta ,\alpha ^{^{\prime }}\beta ^{^{\prime }}}\int_{-\omega }^{0}%
\frac{d\Omega }{2\pi }V_{\mathbf{k}\sigma \alpha \beta }V_{\mathbf{k}\sigma
\alpha ^{^{\prime }}\beta ^{^{\prime }}}\rho _{\beta ^{^{\prime }}\alpha
}\left( \mathbf{k},\Omega \right) \rho _{\beta \alpha ^{^{\prime }}}\left(
\mathbf{k},\Omega +\omega \right) .
\end{equation}

The calculated optical conductivity $\sigma \left( \omega \right) $
is shown in Fig. 14. The black-squared curve shows the result for
the bare five-band itinerant electrons and the red circles represent
the result of the SDW reconstructed bands. Note that the multi-peak
structure is mainly due to the multi-band effect with the lower ones
change significantly in the SDW state. Furthermore, by incorporating
the scattering with the collective spin modes at the RPA level, as
shown by the blue triangles in Fig. 14, no significant change has
been found in the optical conductivity with $\mathbf{q}=0$. It
clearly illustrated that the itinerant electrons remain very
coherent in the AF ordered phase, where the low-lying Goldstone mode
does not strongly scatter the quasiparticles as expected.

\begin{figure}[tbp]
\includegraphics[clip,width=0.7\textwidth]{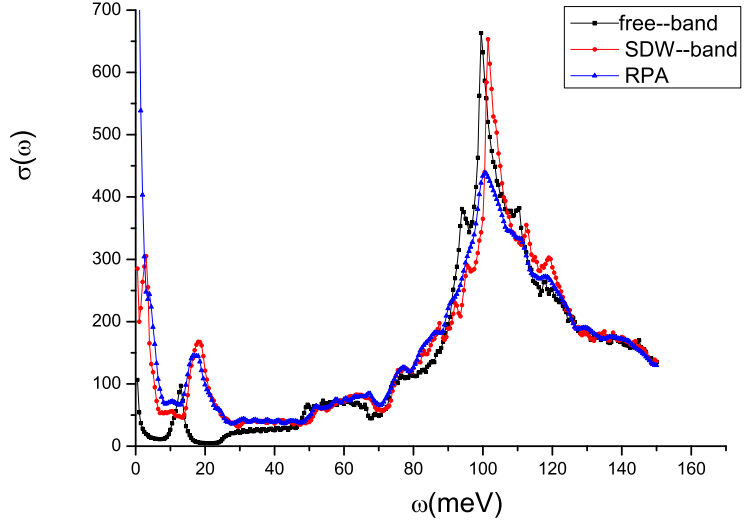}
\caption{(Color online) The optical conductivity $\protect\sigma \left(%
\protect\omega \right)$ in the AF ordered state (blue solid
triangle) which shows that the quasiparticles remain quite coherent
as compared to the case without incorporating the scattering with
the collective spin modes (red solid circles). For comparison,
$\protect\sigma(\protect\omega )$ for the pure five-band model
without the SDW reconstruction is shown (black square).}
\label{Fig.14}
\end{figure}

\section{Discussion and conclusion}

In this work, we have studied the collective spin excitations in the AF
ordered phase of a multi-component system composed of coexistent itinerant
and localized electrons. The main prediction is that a usual spin mode is
split into two branches in such a multi-band system with orbital-selective
Mott transition. The lower branch is a gapless Goldstone mode which is
recovered at the RPA level and is quickly damped above $2\Delta _{\mathrm{SDW%
}}$ by coupling to the particle-hole continuum of itinerant electrons,
similar to the case in which the SDW order is due to the pure Fermi-surface
nesting effect for itinerant electrons. However, an upper branch remnant
spin wave reemerges above the SDW gap over a much wider energy $\sim J_{2}$
which is dominantly contributed by the local moment fluctuations. Here the
lower and upper branches can be regarded as in-phase and out-of-phase
combinations of the spin fluctuations from the itinerant and Mott-localized
electrons, which are clearly distinguished from the single mode in a
conventional AF state either due to the pure Fermi-surface nesting effect
for itinerant electrons or AF superexchanges of local moments.

Experimentally the high-energy spin-wave excitation has been clearly
observed by the neutron scattering experiments over an energy scale $\sim
J_{2}$ and presumably survives in the high-temperature regime above the
ordered phase. However, a small gap ($\sim 6-10$ \textrm{meV}$)$ has been
generally found in \textrm{SrFe}$_{2}$\textrm{As}$_{2}$ and \textrm{BaFe}$%
_{2}$\textrm{As}$_{2}$, and interpreted as due to ion-anisotropy\cite%
{neutron1,neutron2}. As shown in this work, a small ion-anisotropy can
indeed easily destroy the lower branch Goldstone mode, while the upper
branch is more robust. It remains to be seen if the lower branch spin mode
can be unambiguously identified for a sample with less ion-anisotropy.

Another distinct property of the present system is that the itinerant
electrons become very coherent in the AF ordered phase, leading to a good
metallic behavior after the AF transition. This is in contrast to the
presumably strong scattering between the itinerant electrons and local
moments in the normal state, where due to the very fact that the momentum
displacement of the hole-electron Fermi pockets of the itinerant electrons
matches with the AF wavevector of the local moments, there exists a strongly
enhanced interaction, i.e., the \textquotedblleft \emph{resonant
effect\textquotedblright \ }around $\mathbf{Q}_{s}$, between the two
subsystems. It provides the strong scattering source responsible for a
drastic change in the charge response once the system enters the AF
long-range ordered state at low-temperature. By forming a joint magnetic
ordering, the two subsystem effectively get \textquotedblleft
decoupled\textquotedblright \ as the AF fluctuations of the local moments
gain a gap $\eta $ as the out-of-phase collective mode. On the other hand,
the gapless Goldstone mode is effectively decoupled from the itinerant
electrons in the long-wavelength, thanks to the Fermi surface reconstruction
by the SDW order.

Therefore, the collective fluctuations of the local moments can serve as the
main driving force for both the AF ordering as well as the superconducting
pairing in the system via the \textquotedblleft resonant
effect\textquotedblright \ on itinerant electrons, as first pointed out in
Ref. \cite{model}. The resulting magnetic and charge properties in the AF
ordered state, in particular the two branch collective spin modes predicted
in the present work, can be further tested by experiment in order to
establish the relevance of the model with the iron pnictides.

\begin{acknowledgments}
We acknowledge stimulating discussions with X.H. Chen, P.C. Dai,
D.L. Feng, D.H. Lee, P.A. Lee, T. Li, Z.Y. Lu, N.L. Wang, T. Xiang,
Y.Z. You, G.M. Zhang and H. Yao. The authors are grateful for the partial support by
NSFC grant Nos. 10688401, 10704008,  10834003 and 10874017 as well as
the National Program for Basic Research of MOST.
\end{acknowledgments}

\end{document}